\documentclass[conference]{IEEEtran}
\pdfoutput=1
\ifdefined\pdfsuppresswarningpagegroup
\fi

\usepackage{amsmath}
\usepackage{booktabs}
\usepackage{cite}
\usepackage{graphicx}
\usepackage{tabularx}
\usepackage{url}
\usepackage{xcolor}
\usepackage{listings}
\usepackage{tikz}

\newcommand{\ocudu}{OCUDU}
\newcommand{\us}{\ensuremath{\mu\mathrm{s}}}
\newcommand{\code}[1]{\texttt{#1}}
\newcommand{\ClassA}{Class~A}
\newcommand{\ClassB}{Class~B}
\newcommand{\ClassC}{Class~C}
\newcolumntype{Y}{>{\raggedright\arraybackslash}X}
\newcolumntype{L}[1]{>{\raggedright\arraybackslash}p{#1}}

\usetikzlibrary{arrows.meta,positioning,fit,backgrounds,calc,patterns,decorations.pathreplacing}

\definecolor{ocuA}{HTML}{1F6FB2}     
\definecolor{ocuB}{HTML}{D97706}     
\definecolor{ocuC}{HTML}{0E8A6C}     
\definecolor{ocuE}{HTML}{7C3AED}     
\definecolor{ocuH}{HTML}{4B5563}     
\definecolor{ocuP}{HTML}{9CA3AF}     
\definecolor{ocuR}{HTML}{B91C1C}     

\tikzset{
  fbox/.style={draw=#1!75, fill=#1!8, rounded corners=2.5pt, align=center,
               inner sep=3.5pt, font=\scriptsize\sffamily},
  fboxsolid/.style={draw=#1!85, fill=#1!75, text=white, rounded corners=2.5pt,
               align=center, inner sep=3.5pt, font=\scriptsize\sffamily\bfseries},
  flane/.style={draw=#1!50, fill=#1!4, rounded corners=4pt, inner sep=4.5pt},
  flabel/.style={font=\scriptsize\sffamily\bfseries, text=#1!80!black},
  fnote/.style={font=\fontsize{6}{7}\selectfont\sffamily, text=#1!70!black, align=center},
  farr/.style={-{Latex[length=1.8mm]}, line width=0.65pt, draw=#1!85},
  fdash/.style={-{Latex[length=1.8mm]}, dashed, line width=0.65pt, draw=#1!85},
  fbrace/.style={decorate, decoration={brace, amplitude=3pt}, line width=0.6pt, draw=#1!80},
}

\usepackage[hidelinks,breaklinks]{hyperref}

\begin{document}

\title{The \ocudu{} dApp Platform: An Open Runtime and E3 Interface for Real-Time AI-RAN}

\author{
\IEEEauthorblockN{Timothy O'Shea, Matthew Pennybacker, Andriy Kharchenko}
\IEEEauthorblockA{DeepSig Inc., Arlington, VA, USA}
}

\maketitle

\begin{abstract}
Machine learning has shown its largest gains in the band below 10\,ms inside a 3GPP new radio (NR) 5G distributed unit (DU): link adaptation, per-slot scheduling, channel estimation, and the receiver itself. No open platform has let independently built software run there. Prior dApp frameworks reached the band only as external observers of an export stream. This paper is a guided introduction to the \ocudu{} dApp platform, an open runtime and E3 interface under which signed AI-RAN applications execute inside a production DU under three timing contracts: resident on the GPU receive chain (\ClassA{}), inside the scheduler's 100\,\us{} admitted deadline (\ClassB{}), or as never-blocking observers whose results the scheduler consumes (\ClassC{}). The conventional path is never displaced, and every authority is typed, validated, and operator-bounded. The paper explains how the runtime, the embedded E3 agent, and the three public repositories fit together; shows a dApp's source, its signed package, and its lifecycle state machine; defines the contracts a module is written against; and shows how one management surface serves a Python script, an operator's console, and an LLM agent. On a GB10 gNB with attached handsets, dApps of all three classes, including an out-of-tree neural equalizer, ran together on a live cell without a single fallback, and equalizer variants were compared over the air by lifecycle operations alone. Every measured checkpoint is reported with its conditions and its gaps. Platform, SDK, and a zero-hardware quickstart are public under BSD-3-Clause-Clear as a preview release of the \ocudu{} AI-RAN Working Group~2, inviting feedback, new use cases, and independent vetting ahead of upstreaming into the \ocudu{} mainline.
\end{abstract}


\section{Introduction}
\label{sec:intro}

The O-RAN control architecture leaves its most valuable real estate unmanaged. rApps operate at 1\,s and above, and xApps reach down to roughly 10\,ms~\cite{polese_oran,oranwg3}. Everything faster has been sealed inside the vendor's DU: link adaptation, per-slot scheduling, channel estimation, and any reaction to the spectrum inside a coherence time. Yet this band is where learned components have shown the largest gains, from neural receivers~\cite{neural_rx_2024} to learned link adaptation, and it is where sensing applications such as integrated sensing and communication (ISAC) must live to see the channel at all.

\begin{figure*}[t]
  \centering
  \resizebox{0.98\textwidth}{!}{
\begin{tikzpicture}[x=2.05cm]
  \draw[line width=0.8pt, {Latex[length=2mm]}-{Latex[length=2mm]}, draw=ocuH]
    (7.25,0) -- (-0.15,0);
  \foreach \x/\lbl in {0/{1\,$\mu$s},1/{10\,$\mu$s},2/{100\,$\mu$s},3/{1\,ms},4/{10\,ms},5/{100\,ms},6/{1\,s},7/{10\,s}}{
    \draw[draw=ocuH] (\x,-0.06) -- (\x,0.06);
    \node[font=\tiny\sffamily, text=ocuH, below=1pt] at (\x,-0.06) {\lbl};
  }
  \node[font=\scriptsize\sffamily\bfseries, text=ocuH] at (5.5,-0.62) {control-loop latency};

  \fill[ocuH!80] (2.7,0.06) -- (2.62,0.22) -- (2.78,0.22) -- cycle;
  \node[fnote=ocuH, anchor=south] at (2.7,0.24) {1 slot $=$ 500\,$\mu$s (30\,kHz SCS)};

  \node[fbox=ocuH, fit={(5.98,0.62) (7.25,1.17)}, inner sep=0pt,
        label={[flabel=ocuH]center:{rApps · Non-RT}}] {};
  \node[fbox=ocuH, fit={(4.0,0.62) (5.93,1.17)}, inner sep=0pt,
        label={[flabel=ocuH]center:{xApps · Near-RT RIC}}] {};
  \node[fbox=ocuE, fit={(0.0,0.62) (3.95,1.17)}, inner sep=0pt,
        label={[flabel=ocuE]center:{dApps --- co-located with the DU}}] {};
  \node[fnote=ocuE] at (1.97,1.42) {below the RIC's reach: no standardized tier before dApps};

  \node[draw=ocuP!90, pattern=north east lines, pattern color=ocuP!70,
        rounded corners=2.5pt, fit={(2.95,-1.62) (4.0,-1.07)}, inner sep=0pt] {};
  \node[fnote=ocuP!40!black, anchor=west] at (4.1,-1.34)
    {\textbf{prior dApp frameworks:} external observer loop,\\[-1pt] IQ export $+$ coarse indications};

  \node[fboxsolid=ocuC, fit={(2.55,-2.32) (4.0,-1.77)}, inner sep=0pt,
        label={[flabel=white]center:{Class C · async}}] {};
  \node[fboxsolid=ocuB, fit={(1.75,-2.32) (2.45,-1.77)}, inner sep=0pt,
        label={[flabel=white]center:{Class B}}] {};
  \node[fboxsolid=ocuA, fit={(0.45,-2.32) (1.65,-1.77)}, inner sep=0pt,
        label={[flabel=white]center:{Class A · inline L1}}] {};
  \node[fnote=ocuA] at (1.05,-2.62) {slot-synchronous,\\GPU-resident replacement};
  \node[fnote=ocuB] at (2.1,-2.62) {100\,$\mu$s\\bounded intents};
  \node[fnote=ocuC] at (3.28,-2.62) {observe $+$ advise; producer\\never waits; closes the loop};

  \draw[fbrace=ocuE] (4.0,-3.0) -- (0.45,-3.0)
    node[midway, below=5pt, flabel=ocuE]
    {OCUDU dApp platform: one package format, one E3 plane, three timing contracts};
\end{tikzpicture}}
  \caption{The control-loop ladder. O-RAN's RICs stop at about 10\,ms. Prior dApp frameworks entered the sub-10\,ms band only as external observer loops (hatched). \ocudu{} covers the whole band with three timing contracts under one package format and one E3 plane.}
  \label{fig:ladder}
\end{figure*}
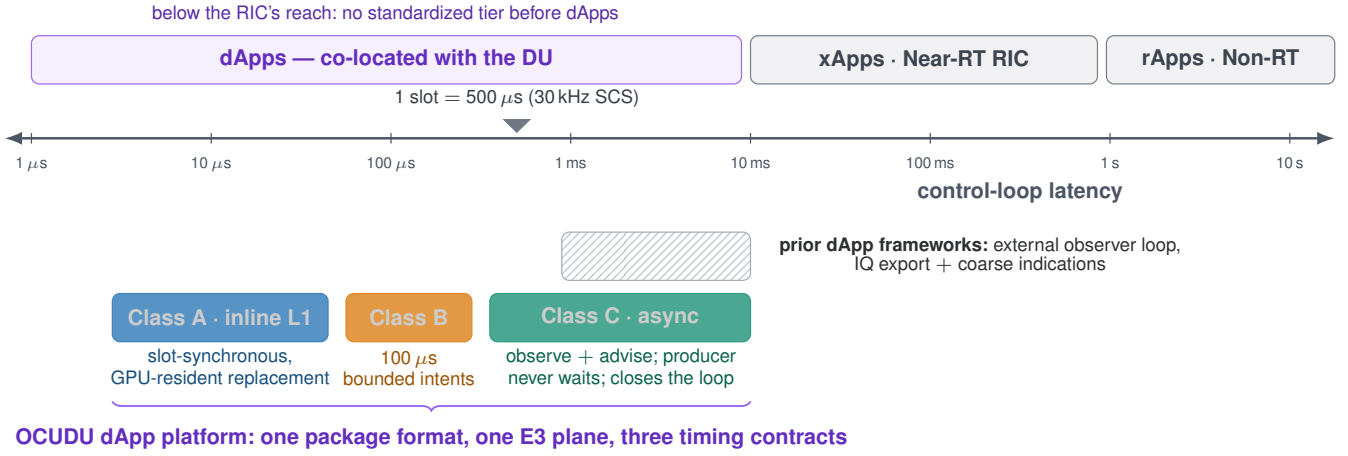

The dApp concept named this tier and proved the demand. Bonati et al.\ proposed it~\cite{bonati2022dapps}; Lacava et al.\ gave it the E3 interface, the real-time interface between a dApp and its host RAN node, and a working OpenAirInterface framework~\cite{lacava2025dapps,openrangymtutorial,libe3}; NVIDIA's Aerial dApp container echoed it~\cite{villa2026aerialdapp,nvidia_aerial,cohenarazi2025ai_aerial}. The O-RAN research arm has since published dApp use cases~\cite{oran_ngrg}, vendors are opening their lower PHY~\cite{nokia2026dapps}, and deployment characterizations are appearing~\cite{boeira2026perf}. These systems share one structural ceiling, shown hatched in Fig.~\ref{fig:ladder}: the dApp is an \emph{external observer loop}. L1 exports IQ or KPIs, a separate process consumes them, and control returns as a coarse indication with no admission bound, no deadline, and no fallback. The highest-value applications, a neural receiver on the L1's own tensors or a learned policy answering before the scheduling decision commits, cannot be expressed that way.

This paper introduces the \ocudu{} dApp platform, which opens the sub-10\,ms band as a managed, open runtime. The platform rests on two claims. First, AI-RAN's upside arrives quickly only if this band is open to a diverse community of vendors, operators, and researchers. Second, one infrastructure can carry everything that community builds. The same package format, E3 plane, and safety machinery that let a Python script watch CRC verdicts also let a neural receiver replace the receive chain, a sensing worker steer the scheduler, a continual-learning loop run from live capture to model rollback, and several vendors ship dApps that one operator composes at runtime.

The paper is written as a tutorial. It proceeds from the architecture at a glance and the vocabulary it uses (Section~\ref{sec:glance}) through the three execution classes and their contracts (Section~\ref{sec:classes}), packages, trust, and the lifecycle state machine (Section~\ref{sec:lifecycle}), the E3 plane (Section~\ref{sec:e3}), where the code lives (Section~\ref{sec:code}), a first session with real source for a \ClassB{} module, a portable \ClassC{} client, and an agent (Section~\ref{sec:using}), the applications it carries (Section~\ref{sec:walkthroughs}), the vendor ecosystem (Section~\ref{sec:ecosystem}), the latency and safety architecture with its fallback semantics (Section~\ref{sec:safety}), the measured evidence with its limits (Section~\ref{sec:evidence}), and what the expanded contract buys (Section~\ref{sec:compare}). The measured interface study behind the class definitions is reported separately~\cite{oshea2026dappstudy}.

\section{Prior dApp Frameworks and Their Boundary}
\label{sec:prior}

The reference dApp implementation~\cite{lacava2025dapps} pairs a Python authoring library with \emph{libe3}~\cite{libe3}, a patched OpenAirInterface gNB exporting IQ through an E3 service model, and a FlexRIC extension for xApp--dApp coordination; its canonical dApp senses the spectrum on exported IQ and returns a PRB blacklist for the MAC scheduler to enforce. NVIDIA's container~\cite{villa2026aerialdapp,nvidia_aerial} pairs an E3 manager with a Triton inference server beside the CUDA L1, publishing RAN data through shared-memory descriptors to GPU inference backends, applied to real-time uplink interference detection in InterfO-RAN~\cite{santhi2025interforan}. Both are genuine sub-RIC control, and both validated E3 as the interface.

Three consequences follow from the external-loop shape regardless of implementation quality. \emph{First}, inline applications cannot be expressed: a neural receiver must run on the L1's tensors and stream inside the slot, and an external process pays serialization, wakeup, and CUDA-context costs to reach data the PHY already holds~\cite{pennybacker2026cudaocudu}. \emph{Second}, the radio inherits the dApp's failure modes: an experimental component can degrade or stall the cell, because there is no producer-side isolation, admission control, deadline, or always-available conventional path. \emph{Third}, operations do not scale: every dApp is a bespoke deployment, because there is no signed packaging, typed configuration, model lifecycle, or uniform observability, and no framework gives an operator the machinery to compose dApps from several vendors on one cell. The platform addresses each while keeping the observer pattern fully supported as its entry point (\ClassC{} over E3), so existing dApp-style consumers port directly.

\section{The Platform at a Glance}
\label{sec:glance}

\begin{figure*}[t]
  \centering
  \resizebox{0.98\textwidth}{!}{
\begin{tikzpicture}[x=1cm, y=1cm, >=Latex,
  stage/.style={draw=ocuH!70, fill=ocuH!6, rounded corners=1.5pt, minimum height=7mm, minimum width=13mm, align=center, inner sep=2pt, font=\scriptsize\sffamily},
  dapp/.style={draw=#1!85, very thick, fill=#1!12, rounded corners=2pt, minimum height=8.5mm, align=center, inner sep=3pt, text width=24mm, font=\scriptsize\sffamily},
  store/.style={draw=ocuH!70, fill=yellow!18, rounded corners=1pt, align=center, inner sep=2.5pt, minimum height=6mm, font=\scriptsize\sffamily},
  proc/.style={draw=black!55, dashed, rounded corners=4pt, inner sep=6pt},
  flow/.style={->, thick, #1!85}, flow/.default=ocuH,
  lbl/.style={font=\tiny\sffamily, fill=white, inner sep=1pt, text=#1!80!black}, lbl/.default=ocuH,
]
\node[stage] (grid) at (0.8,0) {UL grid\\(GPU)};
\node[stage] (ce) at (2.5,0) {CE};
\node[stage] (eq) at (4.0,0) {EQ};
\node[stage] (dem) at (5.5,0) {demap};
\node[stage] (ldpc) at (7.0,0) {LDPC};
\foreach \a/\b in {grid/ce, ce/eq, eq/dem, dem/ldpc} { \draw[->, ocuH!70] (\a) -- (\b); }
\node[dapp=ocuA, text width=30mm] (a) at (4.0,1.8) {\textbf{Class A module}\\CE, CE$+$EQ, or full RX on the lane's CUDA stream; completion event};
\draw[flow=ocuA] (ce.north) |- ($(a.west)+(0,-0.15)$);
\node[lbl=ocuA] at (1.65,1.05) {invoke, shape admitted};
\draw[flow=ocuA] ($(a.east)+(0,-0.15)$) -| (dem.north);
\node[lbl=ocuA] at (6.55,1.05) {DU-owned outputs};
\node[stage, draw=ocuR!60, fill=white, text width=18mm] (fb) at (8.2,1.8) {conventional stage armed as fallback};
\draw[->, thin, ocuR!60, dashed] (a.east) -- (fb.west);
\node[stage, minimum width=22mm] (mac) at (1.3,-1.9) {MAC scheduler\\decision boundary};
\node[dapp=ocuB] (b) at (4.5,-1.9) {\textbf{Class B module}\\direct call, 100\,$\mu$s deadline, allow flags};
\draw[flow=ocuB] ($(mac.east)+(0,0.12)$) -- ($(b.west)+(0,0.12)$); \node[lbl=ocuB] at (2.75,-1.45) {candidates, features};
\draw[flow=ocuB] ($(b.west)+(0,-0.12)$) -- ($(mac.east)+(0,-0.12)$); \node[lbl=ocuB] at (2.75,-2.35) {validated intents};
\node[store, text width=21mm] (ctx) at (7.7,-1.9) {host-owned caches\\interference map, quiet reservations};
\draw[flow=ocuH] (ctx.west) -- (b.east);
\node[stage, minimum width=22mm] (pub) at (1.3,-3.7) {slot publisher\\try-lock $+$ eventfd};
\draw[flow=ocuH] (grid.south) -- (grid.south |- pub.north); \node[lbl] at (0.8,-1.2) {capture};
\node[dapp=ocuC] (cn) at (4.5,-3.7) {\textbf{Class C native}\\leased view, 0 copies};
\draw[flow=ocuC] (pub.east) -- (cn.west); \node[lbl=ocuC] at (2.75,-3.4) {lease};
\draw[flow=ocuC] (cn.east) -| (ctx.south); \node[lbl=ocuC] at (6.5,-3.4) {results};
\node[stage, minimum width=36mm, fill=ocuE!10, draw=ocuE!80] (agent) at (2.0,-5.5) {\textbf{embedded E3 agent}\\lifecycle, config, models, metrics,\\incidents, streams, RAN control};
\node[stage, minimum width=28mm, fill=ocuE!6, draw=ocuE!60] (tp) at (6.3,-5.5) {telemetry publisher\\8 streams, rate floors,\\idle $=$ 1 atomic load};
\draw[->, thin, ocuE!70, dashed] (tp.north) |- (cn.south east);
\draw[->, thin, ocuE!70, dashed] (agent.north) -- (agent.north |- pub.south);
\draw[->, thin, ocuE!70, dashed] (agent.east) -- (tp.west);
\node[lbl=ocuE] at (4.15,-4.7) {manages every class off the hot path};
\begin{scope}[on background layer]
\node[proc, fit=(grid)(ldpc)(a)(fb)(mac)(b)(ctx)(pub)(cn)(agent)(tp), inner ysep=10pt] (du) {};
\end{scope}
\node[font=\small\sffamily\bfseries, text=ocuH, anchor=south west] at (du.north west) {OCUDU DU process (gNB): frozen, size-tagged C ABI v1};
\node[store, fill=orange!18, text width=18mm] (ring) at (11.6,-3.7) {shared-memory ring\\or CUDA-IPC pool};
\node[dapp=ocuC] (cs) at (14.6,-3.7) {\textbf{Class C supervised worker}\\seccomp, heartbeats, restart};
\node[stage, text width=20mm, fill=white] (seq) at (14.6,-5.5) {SEQPACKET control\\credentials checked};
\draw[flow=ocuC] (ring.east) -- (cs.west);
\draw[flow=ocuC] (pub.south east) ++(0,0.1) -- ++(0,-0.35) -| ($(ring.south)+(0,-0.35)$) -- (ring.south);
\node[lbl=ocuC] at (9.6,-4.35) {1 copy, 1 eventfd; full ring $=$ counted drop};
\draw[<->, thin, ocuH!70] (agent.east -| du.east) -- (seq.west); \node[lbl] at (12.2,-5.25) {supervise, restart};
\begin{scope}[on background layer]
\node[proc, fit=(ring)(cs)(seq), inner ysep=9pt] (wk) {};
\end{scope}
\node[font=\small\sffamily\bfseries, text=ocuC, anchor=south west] at (wk.north west) {worker process, same host};
\node[dapp=ocuE, text width=15mm] (cli1) at (11.5,-0.3) {\code{dappctl}\\operations};
\node[dapp=ocuE, text width=15mm] (cli2) at (13.6,-0.3) {\code{ocudu\_e3}\\Python client};
\node[dapp=ocuE, text width=15mm] (cli3) at (15.7,-0.3) {MCP server\\LLM agents};
\node[dapp=ocuE, text width=30mm] (cli4) at (12.55,1.3) {\textbf{portable Class C dApps}\\any language, container, or host};
\node[dapp=ocuE, text width=15mm] (cli5) at (15.7,1.3) {recorders,\\dashboards};
\draw[flow=ocuE] (du.east |- cli1.west) ++(0,0.15) -- ($(10.55,-0.3)+(0,0.15)$);
\draw[flow=ocuE] ($(10.55,-0.3)+(0,-0.15)$) -- ($(du.east |- cli1.west)+(0,-0.15)$);
\node[lbl=ocuE] at (10.05,0.15) {E3AP APER / SCTP 36423};
\node[lbl=ocuE] at (10.05,-0.75) {E3DP FlatBuffers / SCTP 38472};
\node[lbl=ocuE] at (10.05,-1.3) {credentialed local socket};
\begin{scope}[on background layer]
\node[proc, fit=(cli1)(cli2)(cli3)(cli4)(cli5), inner ysep=9pt] (cf) {};
\end{scope}
\node[font=\small\sffamily\bfseries, text=ocuE, anchor=south west] at (cf.north west) {E3 clients: any process, any host};
\end{tikzpicture}}
  \caption{The platform as implemented. Inside the DU process: \ClassA{} modules are invoked on the PUSCH lane's own CUDA stream and write DU-owned outputs; a \ClassB{} module is a direct call at the scheduler's decision boundary returning validated intents; native \ClassC{} modules take leased views of each captured slot and publish results into host-owned caches the scheduler reads back. Outside: supervised \ClassC{} workers behind a shared-memory ring or CUDA-IPC pool, and E3 clients of every kind. One embedded E3 agent manages all of it off the real-time path.}
  \label{fig:platform}
\end{figure*}
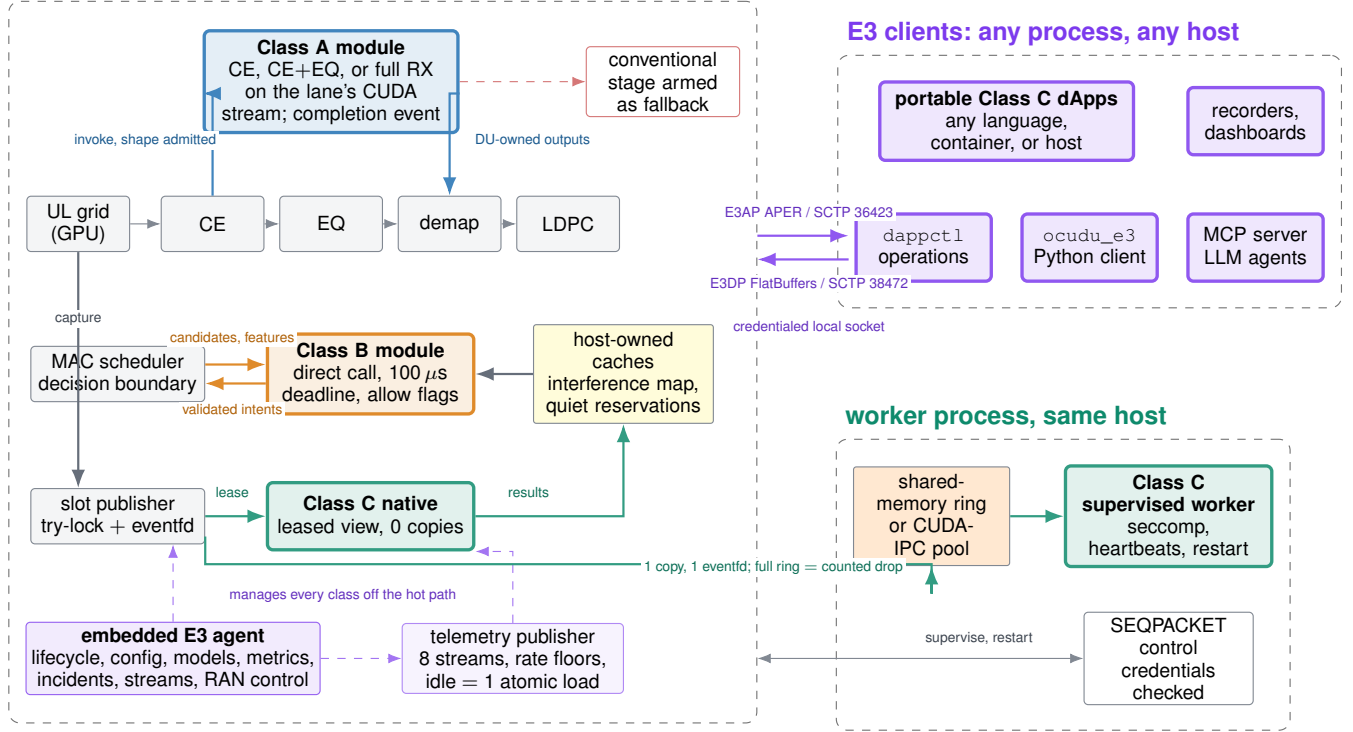

\ocudu{}, the Linux Foundation's open-source 5G CU/DU project~\cite{ocudu_foundation}, is a production-grade 5G NR stack whose GPU-resident CUDA L1 and fronthaul acceleration are described in~\cite{pennybacker2026cudaocudu}. The dApp runtime adds three seams to that stack and one management plane over them (Fig.~\ref{fig:platform}). Table~\ref{tab:glance} states the three seams side by side, and Table~\ref{tab:glossary} defines the terms the rest of the paper relies on. The terms are short, but several of them, \emph{shadow}, \emph{evaluation mode}, \emph{generation}, and \emph{lane}, carry precise meanings that the prose below depends on.

\begin{table*}[!t]
\caption{The three execution classes at a glance. Interface identifiers are the 32-bit contract ids a package declares.}
\label{tab:glance}
\centering
\scriptsize
\begin{tabularx}{\textwidth}{@{}L{1.45cm}L{2.2cm}L{2.25cm}L{2.9cm}L{2.8cm}L{1.5cm}Y@{}}
\toprule
\textbf{Class} & \textbf{Interfaces} & \textbf{Runs} & \textbf{Input} & \textbf{Output} & \textbf{Budget} & \textbf{When it fails or is late} \\
\midrule
\ClassA{}, resident inline L1 & \code{0x00010003} channel estimation; \code{0x00010004} estimation plus equalization; \code{0x00010005} full receiver to soft bits & in the DU process, enqueued on the PUSCH lane's own CUDA stream (CPU variants call the CPU upper PHY directly) & GPU-resident slot grid (complex BF16), DM-RS pilot tensor, compact data-RE indices, typed PUSCH and DM-RS metadata & channel estimates plus noise; equalized symbols plus post-equalization noise; FP16 soft bits before descrambling & 100\,\us{} for estimation, 150\,\us{} to completion for the deeper two & out-of-profile grant, bypass, or a failed enqueue: the conventional stage runs in the same invocation. Late completion: the result is used, an incident is recorded, and eight in a row open the lane breaker \\
\midrule
\ClassB{}, bounded real-time control & \code{0x00010006} scheduler & in the DU process, a direct call at the scheduler's per-slice decision, after the conventional policy has run & up to 8{,}192 pointer-free candidate records, the latest interference map, the free-VRB mask, and the flags the operator allows & intents per candidate: priority or forbid, PRB limits, preference, or assignment, MCS, layers, TPC, and a per-PRB avoid mask & 100\,\us{} from feature construction through commit & too late before the call: skipped. Late or invalid after it: discarded. Late after commit: rolled back. The conventional decision always stands \\
\midrule
\ClassC{}, asynchronous observation and advisory & \code{0x00010001} full uplink spectrum grid; \code{0x00010002} SRS channel estimates & in process on a leased view, in a supervised process on the same host, or as an E3 client on any host & one immutable publication per slot: grid or SRS tensor, the scheduler's expected-use plan, sequence and radio time & per-PRB interference map and quiet-period intents from the one admitted authority package; observations from everyone else & none on the producer; the consumer paces itself & a full ring or a slow consumer is a counted drop; a dead worker is restarted under a bounded budget; the map it published expires by its validity horizon \\
\bottomrule
\end{tabularx}
\end{table*}

\begin{table}[t]
\caption{Vocabulary used in this paper.}
\label{tab:glossary}
\centering
\scriptsize
\begin{tabularx}{\columnwidth}{@{}L{1.9cm}Y@{}}
\toprule
\textbf{Term} & \textbf{Meaning} \\
\midrule
lane, worker & a host thread that invokes a dApp; lane state is lock-free and local, instance state is shared across lanes \\
gate & the single atomic that decides whether a hook is invoked; closed means the conventional path runs \\
bypass & a module declining an invocation; the host runs the conventional stage \\
admission profile & the shape bounds a \ClassA{} interface declares (maximum PRBs, ports, layers, modulations, DM-RS types); a grant outside it takes the conventional path \\
intent & a \ClassB{} module's advice to the scheduler; validated field by field, committed or discarded atomically \\
allow flag & an operator YAML switch granting a \ClassB{} instance authority over one intent class; all default off \\
evaluation mode & YAML \code{scheduler\_evaluation\_only}: a \ClassB{} instance is invoked and validated on every decision, its output discarded and counted \\
shadow & a lifecycle state: warm, configurable, model-loadable, zero traffic exposure; not invoked \\
generation & a monotonic counter on inventory, each instance, each configuration, and each model; mutations carry the expected value and a stale one returns \code{conflict} \\
lease & the bounded lifetime of a \ClassC{} view or a data-store subscription; abandonment lets it lapse \\
hard floor & a per-stream minimum interval no subscriber may go below; non-zero by default \\
breaker & a per-instance latch that opens after eight consecutive faults or misses and routes traffic conventionally until an operator probe \\
incident & an attributed record of a fault, miss, breaker transition, or lifecycle change, in a bounded store \\
preflight & a disposable helper process that queries a module's descriptor before the DU maps it \\
quiescence & proof that no worker is inside a module; required before unload \\
manifest, package sequence & the hash-bound JSON record that is signed, and the monotonic release number folded into the signature \\
trust policy & the operator's offline file of vendor keys, per-package sequence floors, revocations, and a policy epoch \\
quiet reservation & a host-granted, budget-clamped uplink quiet window requested by a sensing authority \\
E3AP, E3DP & the SCTP carriers: APER-encoded management and telemetry on port 36423; a FlatBuffers poll-only data profile on port 38472 \\
\bottomrule
\end{tabularx}
\end{table}

A dApp is a shared object built in the vendor's own repository against the installed SDK, never against the platform's source tree. It is described by a signed manifest and an SPDX software bill of materials (SBOM), staged into a catalog named in the gNB YAML, and driven through an explicit lifecycle over E3 (Section~\ref{sec:lifecycle}). The third-party boundary is a frozen, size-tagged C ABI with a checked-in layout fingerprint verified across GCC and Clang, C11 and C++17, loaded with \code{dlopen}~\cite{posixdlopen}. Cooperation flows only through host-owned, schema-versioned caches. The contracts are backend-neutral: a package declares a CPU (x86, ARM) or CUDA backend and receives matching resident hooks, and the SDK ships every reference in both variants. The emphasis is nonetheless GPU-first, because the compute headroom beside the baseband lets AI-RAN applications be tried at full fidelity, then optimized, without displacing the radio's own budget. Four invariants hold everywhere: the host owns memory, streams, validation, and fallback; the conventional path is never displaced; every authority is typed, validated field by field, and operator-bounded; and dApps never call each other.

\section{The Three Execution Classes}
\label{sec:classes}

\subsection{\ClassA{}: resident inline L1}
\label{sec:classa}
The host supplies GPU-resident tensors and its own per-lane CUDA stream~\cite{cudastreams,cudadriver}. The input is the full-slot device grid in complex BF16, an explicit DM-RS pilot tensor with bounded coordinates, compact ordered data-RE indices, and typed PUSCH metadata (allocation shape, ports, layers, modulation, noise floor, DM-RS layout), so a module can estimate a channel without calling into the private PHY. The outputs are caller-owned device tensors the host prepares and validates: port-and-layer-major channel estimates plus noise at the first depth; compact equalized symbols in \code{[data\_re, layer]} order plus post-equalization noise at the second; FP16 soft bits in \code{[data\_re, layer, bit]} order before descrambling at the third. The three depths rejoin the chain at successively deeper points, and no hook replaces equalization alone, because the equalizer interface also implements channel estimation. A module that succeeds at the second or third depth also supplies the scheduler's uplink SINR from its own post-equalization noise, because the conventional measurement kernels are skipped on that path.

Before any module code runs, the host checks the grant against the interface's declared admission profile with a fixed sequence of integer bounds. The module then enqueues kernels and returns without allocating or synchronizing. Two clocks govern what happens next, and the distinction matters for operators. The \emph{callback clock} covers validation and enqueue; a bypass, an unsupported shape, or a failed enqueue on that clock selects the conventional stage, which is ordered on the same stream \emph{before} the module's writes would land, so fallback is deterministic without host synchronization. The \emph{completion clock} is measured by per-lane CUDA events bracketing the module and its continuation to the pre-descrambling boundary. A late completion is not replayed: the host has already committed to consuming the module's result for that grant, so the result is used, an incident with the observed latency is recorded, a counter increments, and eight consecutive misses open the lane's breaker so later grants take the conventional path. Host-side waiting is bounded by a soft wall budget on the order of 2\,ms, so a wedged kernel costs at most the current slot. The production budgets are 100\,\us{} for estimation and 150\,\us{} to completion for the deeper two depths.

Model management is first class: a bounded stage, validate, warm, activate, rollback sequence swaps weights on a live cell under generation checks with no allocation on the hot path. The reference receiver owns exactly two preallocated device weight banks, publishes the active one with a single atomic exchange, keeps the displaced bank as the only rollback generation, and reuses a bank only after per-lane completion fences prove it drained.

\subsection{\ClassB{}: bounded real-time control}
\label{sec:classb}
The MAC scheduler's per-slice policy computes its conventional priorities first and keeps its normal history. It then builds a pointer-free candidate snapshot and calls the admitted \ClassB{} module directly on its own lane. Table~\ref{tab:classb} lists the complete contract as the header defines it: what each candidate record carries, what an intent may say, and which flags grant which authority. The complete boundary, from feature construction through validation and commit, runs under an admitted deadline of 100\,\us{}, a host default that is deliberately not a YAML knob. Overruns are handled at three checkpoints: too late before the call, the call is skipped; late or invalid after it, the result is discarded; late after commit, everything applied is rolled back. In every case the conventional decision stands, and repeated misses open the instance's breaker.

Authority is granted flag by flag. An unset flag makes that intent class silently unavailable, and the module is told so through \code{allowed\_output\_flags}. The reference scheduler is deliberately a baseline replica whose five stages (ranking, outer-loop link adaptation, MCS selection, PRB packing, uplink power control) all ship off, so activating it with every flag set changes nothing until a stage is switched on over E3. The recommended path is evaluation mode first: the module is invoked and validated on every decision, its output is discarded and counted, and its phase timers accumulate under real load. Only then does the operator flip individual flags.

\begin{table}[t]
\caption{The \ClassB{} contract (\code{use\_cases/v1/scheduler.h}).}
\label{tab:classb}
\centering
\scriptsize
\begin{tabularx}{\columnwidth}{@{}L{1.85cm}Y@{}}
\toprule
\textbf{Element} & \textbf{Content} \\
\midrule
Invocation & direction, slice, radio time, cell PRB count, remaining slice budget, candidate generation, allowed output flags \\
Candidate record (one per UE, at most 8{,}192) & candidate id, RNTI, pending bytes and pending SR, head-of-line delay (downlink), served-rate average, the conventional priority and forbid, CQI or SINR, first-transmission BLER over a 64-outcome window with its count, normalized power headroom, the host's OLLA offset, last outcome and the MCS it used, MCS table and expert limits, layers, transform precoding, target SINR; validity flags per field \\
Context & the latest \ClassC{} interference map as a per-PRB tensor with its provenance header; the free-VRB mask \\
Intent (per candidate) & priority or forbid; minimum and maximum PRBs; preferred start and count (a soft preference, or a hard assignment when allowed); recommended and maximum MCS; layers; TPC command; beam id (placeholder, no v1 authority) \\
Output & the intent array, a decision generation that must match the input, and an optional per-PRB avoid mask for this slot's new uplink transmissions \\
Allow flags (YAML) & \code{avoid\_prb\_mask} (needs a nonzero PRB budget), \code{mcs} (bounded by a maximum delta from the host), \code{prb\_limits}, \code{prb\_preference}, \code{prb\_assignment} (needs the free-VRB snapshot), \code{tpc}, \code{layers} \\
Validation & every candidate covered exactly once; every flagged field in range and permitted; storage identities unchanged; all or nothing \\
\bottomrule
\end{tabularx}
\end{table}

\subsection{\ClassC{}: asynchronous observation and advisory}
\label{sec:classc}
\ClassC{} is where prior architectures ended and where \ocudu{} deliberately begins. A PHY producer creates one immutable publication per slot: the full uplink grid or the SRS channel estimates, the scheduler's committed expected-use plan (a symbol-by-PRB category map, so a radiometer can tell planned-empty cells from our own UEs' energy), a sequence number, and radio time. Three isolation levels consume it: native in-process modules over leased zero-copy views or route-owned snapshots; supervised worker processes over a shared-memory ring, or over a bounded CUDA-IPC export pool filled by one device-to-device snapshot; and portable E3 clients on any host. Nothing ever maps live L1 memory into another process. The producer never waits: publication is a try-lock plus a non-blocking eventfd, one \code{write} and two \code{read} syscalls, and a full ring or busy consumer is a counted drop. Host-inline \ClassC{} copies the full grid on the uplink PHY thread, roughly 50 to 100\,\us{} per slot for a 273-PRB, four-port grid, so multi-cell deployments use the CUDA route, which costs one device-to-device copy instead.

\begin{figure*}[t]
  \centering
  \resizebox{0.98\textwidth}{!}{
\begin{tikzpicture}[x=1cm, y=1cm, >=Latex,
  lane/.style={font=\small\sffamily\bfseries, text=#1, anchor=east},
  solid/.style={draw=#1!85, fill=#1!80, text=white, rounded corners=2pt, align=center, inner sep=3pt, font=\scriptsize\sffamily\bfseries, minimum height=7mm},
  soft/.style={draw=#1!70, fill=#1!8, rounded corners=2pt, align=center, inner sep=3pt, font=\scriptsize\sffamily, minimum height=7mm},
  note/.style={font=\scriptsize\sffamily, text=#1!80!black, align=left},
  arr/.style={->, thick, #1!85},
]
\foreach \i in {0,...,13}{ \draw[draw=ocuH!60, fill=ocuH!12] ({\i*6.5/14},4.55) rectangle ({(\i+1)*6.5/14 - 0.05},4.95); }
\node[note=ocuH, anchor=south west] at (0,5.0) {14 OFDM symbols of uplink slot $n$ arrive (500\,$\mu$s at 30\,kHz SCS)};
\draw[dashed, line width=0.6pt, draw=ocuH!70] (6.5,-0.1) -- (6.5,4.95);
\node[note=ocuH, anchor=north] at (6.5,-0.15) {slot $n$ ends};
\draw[line width=0.7pt, ->, draw=ocuH] (0,0.1) -- (14.6,0.1);
\node[note=ocuH, anchor=north] at (10.8,-0.05) {slots $n{+}1$, $n{+}2$, \dots};
\node[lane=ocuC] at (-0.2,3.9) {Class C};
\node[solid=ocuC, anchor=west] (csnap) at (5.6,3.9) {snapshot $+$ publish};
\node[note=ocuC, anchor=east, align=right] at (5.45,3.9) {try-lock, 1 device copy,\\1 eventfd write; a full ring\\is a counted drop, never a wait};
\node[soft=ocuC, anchor=west, text width=46mm] (cproc) at (9.0,3.9) {consumer analyzes at its own pace: native module, supervised worker, or E3 client};
\draw[arr=ocuC] (csnap) -- (cproc);
\node[lane=ocuB] at (-0.2,2.55) {Class B};
\node[soft=ocuH, anchor=west] (bconv) at (0.4,2.55) {conventional priorities\\(always computed first)};
\node[solid=ocuB, anchor=west] (binv) at (3.4,2.55) {invoke, validate, commit};
\draw[decorate, decoration={brace, amplitude=3pt}, ocuB!80] (3.4,2.98) -- (binv.east |- 0,2.98) node[midway, above=3pt, note=ocuB] {100\,$\mu$s admitted deadline};
\node[note=ocuR, anchor=west] at (6.7,2.55) {late or invalid: discarded,\\conventional result stands};
\node[soft=ocuB, anchor=west, text width=30mm] (bnext) at (11.6,2.55) {a later decision consumes the map or the quiet grant};
\draw[arr=ocuC] (cproc.south -| 12.4,0) -- (bnext.north -| 12.4,0);
\node[note=ocuC, anchor=west] at (12.55,3.25) {provenance-checked result};
\node[lane=ocuA] at (-0.2,1.3) {Class A};
\node[solid=ocuA, anchor=west] (aker) at (5.3,1.3) {CE / EQ / RX kernels on the host's CUDA stream};
\node[soft=ocuH, anchor=west] (afall) at (5.3,0.55) {conventional receiver armed as per-invocation fallback};
\node[note=ocuA, anchor=east, align=right] at (5.15,1.3) {enqueue and return in microseconds;\\completion clocked by a CUDA event};
\node[note=ocuR, anchor=west] at (12.0,1.3) {budget miss: incident\\$+$ circuit breaker};
\draw[arr=ocuR] (aker.east) -- (11.95,1.3);
\end{tikzpicture}}
  \caption{One uplink slot across the three timing contracts. \ClassB{} decides early inside a bounded boundary; \ClassA{} replaces receive stages inline as the slot completes, measured on the completion clock; \ClassC{} forks a snapshot without blocking and closes the loop at a later decision through provenance-checked results.}
  \label{fig:slot}
\end{figure*}
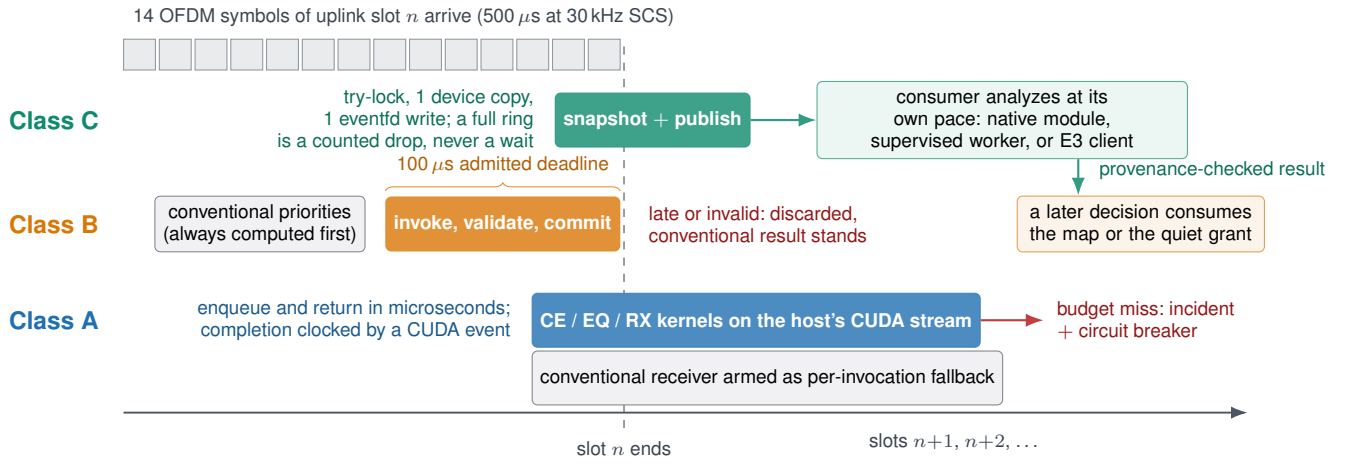

Results flow \emph{back}, but under a strict authority model. A package's manifest carries a capability bit that distinguishes the single admitted map \emph{authority} from any number of observer-only workers. Only the authority may publish per-PRB interference maps and quiet-period intents, and only through the in-process host API or the supervised worker's authenticated control channel, whose peer credentials are pinned at spawn. Each result must match a single-use ledger entry recorded when its input was committed, and the host, not the worker, stamps instance identity, generation, and sequence. A result frame from an observer package is a fatal capability violation. E3 clients cannot publish maps at all; they act back only through management operations. The scheduler's avoidance mask consumes the map and any granted quiet reservation on a later decision, typically within a few slots, and every map carries a validity horizon after which it is ignored. Fig.~\ref{fig:slot} places all three contracts on one slot.

\subsection{A continuum of performance and trust}
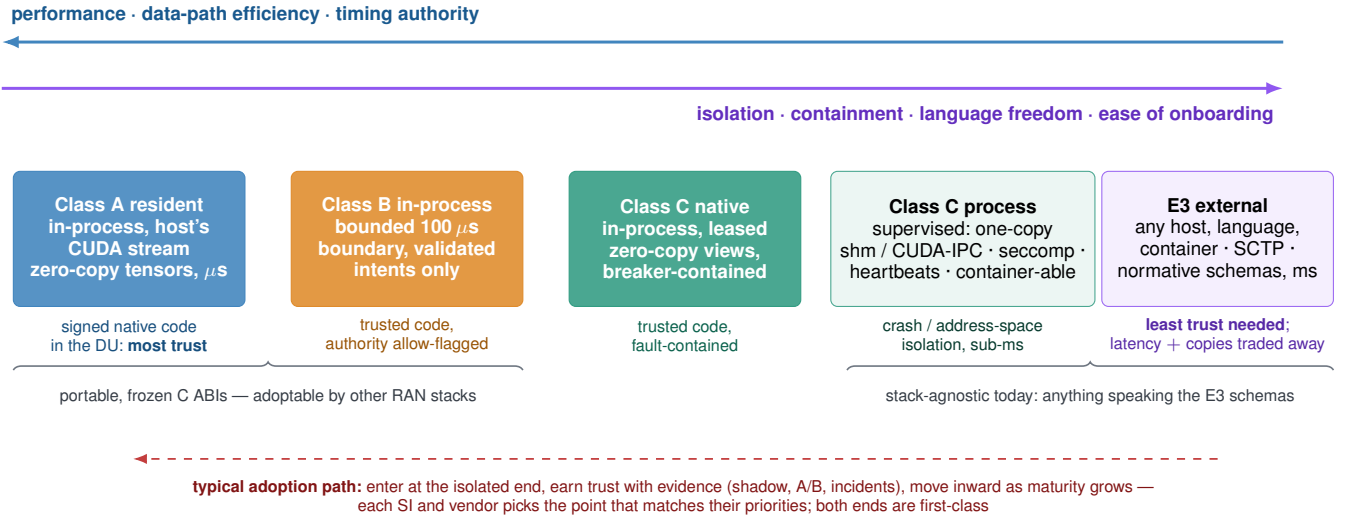
\begin{figure*}[t]
  \centering
  \resizebox{0.98\textwidth}{!}{
\begin{tikzpicture}
  \draw[{Latex[length=2.2mm]}-, line width=1.1pt, draw=ocuA!85] (0.3,5.15) -- (16.9,5.15);
  \node[flabel=ocuA, anchor=west] at (0.3,5.5)
    {performance · data-path efficiency · timing authority};
  \draw[-{Latex[length=2.2mm]}, line width=1.1pt, draw=ocuE!85] (0.3,4.55) -- (16.9,4.55);
  \node[flabel=ocuE, anchor=east] at (16.9,4.2)
    {isolation · containment · language freedom · ease of onboarding};

  \node[fboxsolid=ocuA, minimum width=3.0cm, minimum height=1.75cm] (s1) at (1.95,2.6)
    {\textbf{Class A resident}\\in-process, host's\\CUDA stream\\zero-copy tensors, $\mu$s};
  \node[fboxsolid=ocuB, minimum width=3.0cm, minimum height=1.75cm] (s2) at (5.55,2.6)
    {\textbf{Class B in-process}\\bounded 100\,$\mu$s\\boundary, validated\\intents only};
  \node[fboxsolid=ocuC, minimum width=3.0cm, minimum height=1.75cm] (s3) at (9.15,2.6)
    {\textbf{Class C native}\\in-process, leased\\zero-copy views,\\breaker-contained};
  \node[fbox=ocuC, minimum width=3.0cm, minimum height=1.75cm] (s4) at (12.75,2.6)
    {\textbf{Class C process}\\supervised: one-copy\\shm / CUDA-IPC · seccomp ·\\heartbeats · container-able};
  \node[fbox=ocuE, minimum width=3.0cm, minimum height=1.75cm] (s5) at (16.05,2.6)
    {\textbf{E3 external}\\any host, language,\\container · SCTP ·\\normative schemas, ms};

  \node[fnote=ocuA] at (1.95,1.35) {signed native code\\in the DU: \textbf{most trust}};
  \node[fnote=ocuB] at (5.55,1.35) {trusted code,\\authority allow-flagged};
  \node[fnote=ocuC] at (9.15,1.35) {trusted code,\\fault-contained};
  \node[fnote=ocuC!60!black] at (12.75,1.35) {crash / address-space\\isolation, sub-ms};
  \node[fnote=ocuE] at (16.05,1.35) {\textbf{least trust needed};\\latency $+$ copies traded away};

  \node[fnote=ocuH, align=center] at (3.75,0.55)
    {portable, frozen C ABIs --- adoptable by other RAN stacks};
  \node[fnote=ocuH, align=center] at (14.4,0.55)
    {stack-agnostic today: anything speaking the E3 schemas};
  \draw[fbrace=ocuH] (0.45,0.9) -- (7.05,0.9);
  \draw[fbrace=ocuH] (11.25,0.9) -- (17.55,0.9);

  \draw[fdash=ocuR] (16.05,-0.25) -- (2.0,-0.25);
  \node[fnote=ocuR, anchor=north] at (9.0,-0.4)
    {\textbf{typical adoption path:} enter at the isolated end, earn trust with evidence (shadow, A/B, incidents), move inward as maturity grows ---\\each SI and vendor picks the point that matches their priorities; both ends are first-class};
\end{tikzpicture}}
  \caption{The performance--trust continuum. Five placement points span resident in-process execution (maximum authority, maximum required trust) to external E3 processes (maximum isolation and language freedom). The in-process contracts are portable C ABIs; the E3 end is stack-agnostic.}
  \label{fig:continuum}
\end{figure*}

The classes are points on a performance--trust continuum that the design spans on purpose (Fig.~\ref{fig:continuum}). A resident \ClassA{} module runs in process on the host's stream with zero-copy tensors and microsecond authority, and must therefore be the most trusted artifact in the system: signed native code in the DU's address space. An E3 client at the other end runs in any process, language, or host, needs almost no trust, and pays for that isolation in latency, copies, and rate floors. Between them sit \ClassB{}'s bounded in-process boundary, native \ClassC{} placement, and the supervised process with its one-copy export and seccomp containment. A vendor or integrator enters at the placement that matches its containment requirements and maturity, typically the isolated end, and moves inward as evaluation-mode evidence, A/B windows, and incident-free operation accumulate. Placement is a \ClassC{} choice: \ClassA{} and committing \ClassB{} are in process by definition, and a package declaring an out-of-process inline hook is rejected at catalog time. Section~\ref{sec:evidence} reports what each position costs on the reference host, and the companion interface study~\cite{oshea2026dappstudy} measures every position against the mechanisms of the prior frameworks.

\section{Packages, Trust, and the Lifecycle}
\label{sec:lifecycle}

\textbf{What a package is.} A production bundle is four files: the shared object, a JSON manifest, an SPDX~2.3 SBOM, and a detached signature sidecar. The manifest names the package identity, the interfaces, backends, and placements it declares, the SHA-256 of the artifact, and the SHA-256 of the SBOM. One signature covers all of it: the vendor signs a domain-separated digest of the manifest concatenated with the package sequence number, using ECDSA P-256 with SHA-256. Verifying the signature and re-hashing the artifact and SBOM against the manifest therefore authenticates the whole bundle, and folding the sequence in gives rollback protection. The SBOM is checked, not decorative: it must describe exactly one package and one file whose name, version, supplier, and digest equal the manifest's.

\textbf{What the operator holds.} An offline trust policy names every trusted vendor public key by the hash of its SPKI encoding, a per-package sequence floor, a revocation list, and a monotonic policy epoch that the gNB YAML pins with a minimum. Rotating a key adds the new key and bumps the epoch; revoking one invalidates every signature it ever made; blocking a bad build raises that package's floor. The policy file itself is protected by filesystem permissions and the YAML epoch, not by a signature, which is stated plainly in the security model. Admission then binds the artifact hash and SBOM identity, runs descriptor discovery in a disposable preflight process, checks the manifest against the compiled descriptor, and rejects the package before the DU maps it if any of that fails.

\begin{figure*}[t]
  \centering
  \resizebox{0.98\textwidth}{!}{
\begin{tikzpicture}[x=1cm, y=1cm, >=Latex,
  st/.style={draw=#1!85, fill=#1!10, rounded corners=2.5pt, align=center, inner sep=3pt,
             minimum height=7.5mm, minimum width=17mm, font=\scriptsize\sffamily\bfseries, text=#1!60!black},
  side/.style={draw=ocuR!80, fill=ocuR!8, rounded corners=2.5pt, align=center, inner sep=3pt,
             minimum height=7mm, font=\scriptsize\sffamily, text=ocuR!70!black},
  act/.style={font=\scriptsize\sffamily, text=ocuE!80!black, fill=white, inner sep=1pt},
  note/.style={font=\scriptsize\sffamily, align=center, text=#1!75!black},
  arr/.style={->, thick, draw=#1!85},
]
\node[st=ocuH] (disc) at (0,0) {DISCOVERED};
\node[st=ocuH] (load) at (2.55,0) {LOADED};
\node[st=ocuE] (prep) at (5.1,0) {PREPARED};
\node[st=ocuE] (arm) at (7.65,0) {ARMED};
\node[st=ocuC] (shad) at (10.2,0) {SHADOW};
\node[st=ocuA] (act) at (12.75,0) {ACTIVE};
\node[st=ocuH] (drain) at (15.3,0) {DRAINING};
\node[st=ocuH] (inact) at (15.3,-1.75) {INACTIVE};
\node[st=ocuH] (unl) at (12.75,-1.75) {UNLOADED};
\draw[arr=ocuE] (disc) -- node[act, above] {load} (load);
\draw[arr=ocuE] (load) -- node[act, above] {prepare} (prep);
\draw[arr=ocuE] (prep) -- node[act, above] {arm} (arm);
\draw[arr=ocuE] (arm) -- node[act, above] {enterShadow} (shad);
\draw[arr=ocuE] (shad) -- node[act, above] {activate} (act);
\draw[arr=ocuE] (act) -- node[act, above] {deactivate} (drain);
\draw[arr=ocuH] (drain) -- node[act, right] {gate closed, lanes detached} (inact);
\draw[arr=ocuE] (inact) -- node[act, above] {unload} (unl);
\draw[arr=ocuE, rounded corners=3pt] (inact.south) |- (5.1,-2.55) -- (prep.south);
\node[act] at (9.2,-2.55) {prepare again: re-arm the same instance};
\node[note=ocuH, anchor=north] at (disc.south) {in the catalog,\\verified, not mapped};
\node[note=ocuH, anchor=north] at (load.south) {preflighted,\\\code{dlopen}'d; not callable};
\node[note=ocuE, anchor=north] at (prep.south) {all allocation done;\\weights uploaded};
\node[note=ocuE, anchor=north] at (arm.south) {hooks bound, gate\\disabled; warm-up done};
\node[note=ocuC, anchor=north] at (shad.south) {zero traffic; config\\and model ops legal};
\node[note=ocuA, anchor=north] at (act.south) {gate open; the radio\\sees the dApp};
\node[side, text width=36mm] (deg) at (2.2,-3.55) {\textbf{degraded}: a lane failed to bind or a worker died; the conventional path is in use};
\node[side, text width=40mm] (quar) at (7.1,-3.55) {\textbf{quarantined-resident}: quiescence could not be proven; image held, never freed under a live kernel; \code{retryQuiescence}};
\draw[arr=ocuR, dashed] (act.south) .. controls (12.75,-1.2) and (5.5,-2.0) .. (deg.north east);
\draw[arr=ocuR, dashed] (drain.south east) .. controls (16.3,-1.2) and (11.0,-2.4) .. (quar.north east);
\node[note=ocuR, anchor=west, align=left, text width=66mm] at (10.3,-3.55)
  {\textbf{circuit breaker} (per instance, across lanes) is \emph{not} a lifecycle state: eight consecutive faults or misses open it, traffic routes conventionally, \code{health} shows it, \code{breakerProbe} closes it. \textbf{Evaluation mode} (Class~B, YAML) is ACTIVE with the output discarded.};
\node[note=ocuE, anchor=north west, align=left] at (0,-4.55)
  {every transition is an asynchronous job carrying a transaction id and the expected instance generation; a stale generation returns \code{conflict}; every action accepts \code{dryRun}};
\end{tikzpicture}}
  \caption{The instance lifecycle as implemented. E3 actions (violet) drive the transitions; each state's meaning is written under it. The circuit breaker and \ClassB{} evaluation mode are orthogonal to the state and are shown as notes. Every transition is an asynchronous, generation-checked, dry-runnable job.}
  \label{fig:lifecycle}
\end{figure*}
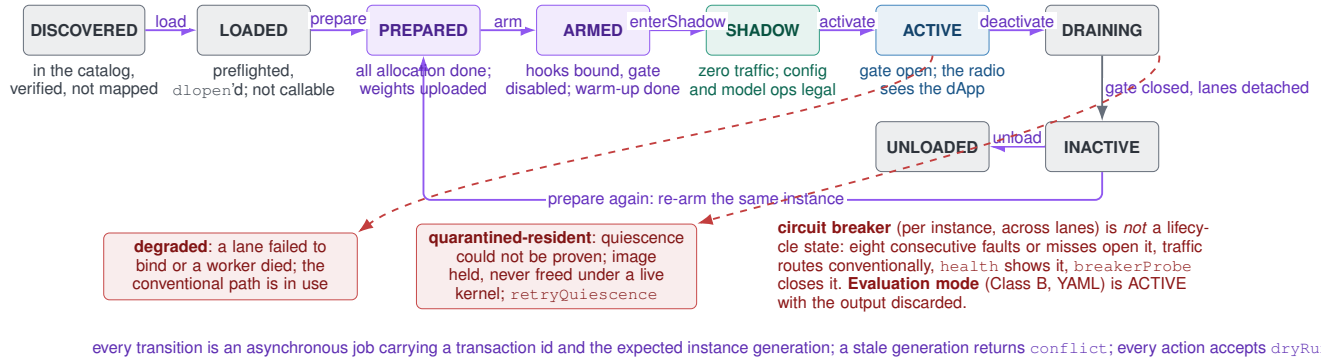

\textbf{The lifecycle.} Fig.~\ref{fig:lifecycle} draws the instance state machine as the runtime implements it. Three states are easy to confuse. \code{LOADED} means preflighted and mapped, not callable. \code{ARMED} means every hook is bound behind a disabled gate and all cold work, including warm-up at the declared worst shape, is done. \code{SHADOW} means the instance is warm, accepts configuration and model operations, and sees zero traffic. Shadow is therefore where weights are staged, validated, and warmed with no exposure, and where an A/B candidate waits; it is not where behaviour is measured. Measurement without authority is evaluation mode for \ClassB{}, which is \code{ACTIVE} with the output discarded, and for \ClassA{} it is activation itself, compared across windows. Activation resolves the declared interface set against the hook registry, stages every lane behind one instance-unique gate, and exposes the whole set with a single release store. Deactivation reverses the order: the gate closes, lanes unpublish after a bounded reader grace period, and only then is the module asked to stop and quiesce. A module that cannot prove quiescence is quarantined resident rather than freed under a live kernel.

\textbf{Concurrency and models.} Inventory, each instance, each configuration, and each model carry a monotonic generation. Every mutating request carries a transaction id, the expected generation, and an optional dry-run flag; a stale generation returns \code{conflict} instead of overwriting, and long operations return a job id instead of holding a socket. Model operations follow a fixed ladder, stage, validate, warm, activate, rollback, retire, over a three-bank store (staged, active, previous) that the SDK implements for every class, so a vendor supplies only the meaning of validate and warm for its own artifact.

\section{The E3 Plane}
\label{sec:e3}

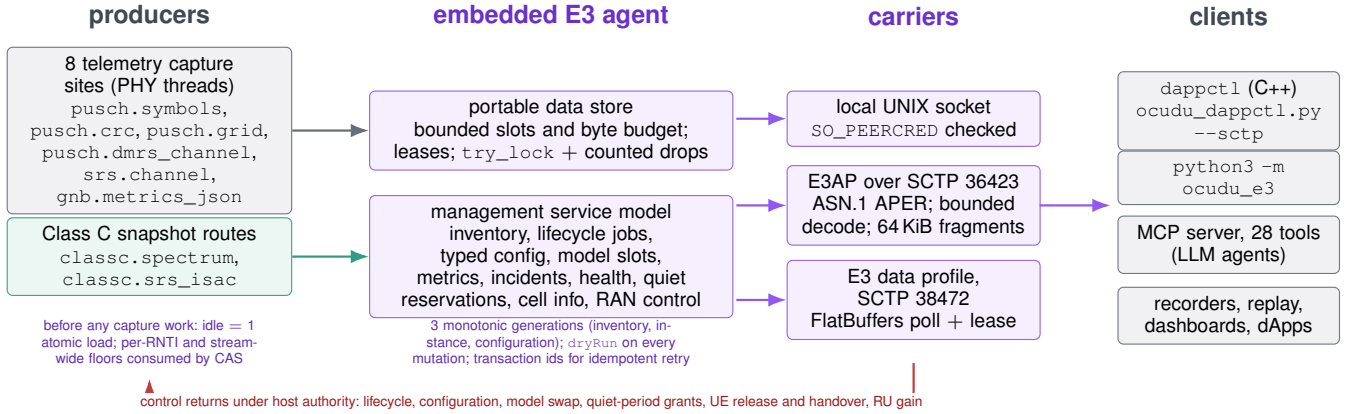
\begin{figure*}[t]
  \centering
  \resizebox{0.98\textwidth}{!}{
\begin{tikzpicture}[x=1cm, y=1cm, >=Latex,
  box/.style={draw=#1!80, fill=#1!8, rounded corners=2pt, align=center, inner sep=3pt, font=\scriptsize\sffamily, text width=34mm},
  head/.style={font=\small\sffamily\bfseries, text=#1!85!black},
  note/.style={font=\tiny\sffamily, text=#1!75!black, align=center, text width=34mm},
  arr/.style={->, thick, #1!85},
]
\node[head=ocuH] at (1.9,5.35) {producers};
\node[box=ocuH] (p1) at (1.9,3.9) {8 telemetry capture sites (PHY threads)\\\code{pusch.symbols}, \code{pusch.crc}, \code{pusch.grid}, \code{pusch.dmrs\_channel}, \code{srs.channel}, \code{gnb.metrics\_json}};
\node[box=ocuC] (p2) at (1.9,2.3) {Class C snapshot routes\\\code{classc.spectrum}, \code{classc.srs\_isac}};
\node[note=ocuE] at (1.9,1.2) {before any capture work: idle $=$ 1 atomic load; per-RNTI and stream-wide floors consumed by CAS};
\node[head=ocuE] at (7.0,5.35) {embedded E3 agent};
\node[box=ocuE, text width=44mm] (store) at (7.0,3.9) {portable data store\\bounded slots and byte budget; leases; \code{try\_lock} $+$ counted drops};
\node[box=ocuE, text width=44mm] (mgmt) at (7.0,2.3) {management service model\\inventory, lifecycle jobs, typed config, model slots, metrics, incidents, health, quiet reservations, cell info, RAN control};
\node[note=ocuE, text width=44mm] at (7.0,1.2) {3 monotonic generations (inventory, instance, configuration); \code{dryRun} on every mutation; transaction ids for idempotent retry};
\draw[arr=ocuH] (p1.east) -- (store.west |- p1.east);
\draw[arr=ocuC] (p2.east) -- (store.west |- p2.east);
\node[head=ocuE] at (11.6,5.35) {carriers};
\node[box=ocuE, text width=30mm] (uds) at (11.6,4.05) {local UNIX socket\\\code{SO\_PEERCRED} checked};
\node[box=ocuE, text width=30mm] (e3ap) at (11.6,2.95) {E3AP over SCTP 36423\\ASN.1 APER; bounded decode; 64\,KiB fragments};
\node[box=ocuE, text width=30mm] (e3dp) at (11.6,1.75) {E3 data profile, SCTP 38472\\FlatBuffers poll $+$ lease};
\draw[arr=ocuE] (9.35,4.05) -- (uds.west);
\draw[arr=ocuE] (9.35,2.95) -- (e3ap.west);
\draw[arr=ocuE] (9.35,1.75) -- (e3dp.west);
\node[head=ocuH] at (15.6,5.35) {clients};
\node[box=ocuH, text width=26mm] (c1) at (15.6,4.15) {\code{dappctl} (C++)\\\code{ocudu\_dappctl.py --sctp}};
\node[box=ocuH, text width=26mm] (c2) at (15.6,3.3) {\code{python3 -m ocudu\_e3}};
\node[box=ocuH, text width=26mm] (c3) at (15.6,2.45) {MCP server, 28 tools\\(LLM agents)};
\node[box=ocuH, text width=26mm] (c4) at (15.6,1.55) {recorders, replay,\\dashboards, dApps};
\draw[arr=ocuE] (13.2,2.95) -- (14.1,2.95);
\draw[arr=ocuR, rounded corners=4pt] (11.6,0.95) -- (11.6,0.45) -- (1.9,0.45) -- (1.9,0.75);
\node[font=\tiny\sffamily, text=ocuR!85!black, fill=white, inner sep=1pt] at (6.75,0.45) {control returns under host authority: lifecycle, configuration, model swap, quiet-period grants, UE release and handover, RU gain};
\end{tikzpicture}}
  \caption{The E3 plane. Producers pay one atomic load per event when nobody listens and consume rate floors by CAS before any capture work. One embedded agent with a bounded store serves three carriers; control returns under host authority.}
  \label{fig:e3}
\end{figure*}

Everything is controlled and observed over E3 (Fig.~\ref{fig:e3}) through published, versioned schemas: an ASN.1 management service model and an E3AP envelope encoded as APER, a telemetry service model, and a FlatBuffers data profile. Three carriers serve them. A credentialed local UNIX socket, mode 0600 with peer credentials checked, serves same-host tooling. E3AP over SCTP on port 36423 carries setup, subscriptions, indications, and management control with message-boundary framing and bounded fragmentation up to 16\,MiB. The data profile (E3DP) over SCTP on port 38472 is a FlatBuffers poll-and-lease endpoint for consumers that do not want an APER stack. The APER decode path is defensively bounded end to end (constraint checks before field reads, bounded reassembly, size-capped strings), so a malformed peer cannot destabilize the agent.

E3 here is a real-time local evolution of E2's setup, subscription, indication, and control philosophy, bridgeable to E2 but not a ratified O-RAN interface. Table~\ref{tab:compat} states exactly how it relates to the two prior implementations. The procedure names and field vocabulary were kept from libe3 so that an adapter is mechanical, but the wire formats are not byte-compatible, and a libe3 dApp needs an envelope re-encode plus a channel-topology shim rather than a direct connection. SCTP is the carrier E2 standardized and the one several large vendors prefer for RAN control planes, which is why it is a first-class carrier here.

\begin{table}[t]
\caption{Wire compatibility with the prior E3 implementations.}
\label{tab:compat}
\centering
\scriptsize
\begin{tabularx}{\columnwidth}{@{}L{1.6cm}L{2.1cm}L{2.0cm}Y@{}}
\toprule
 & \textbf{\ocudu{} E3} & \textbf{libe3 (NEU)} & \textbf{NVIDIA cuSense} \\
\midrule
Encoding & ASN.1 APER & ASN.1 APER or JSON & JSON \\
Transport & SCTP with message boundaries & ZMQ or POSIX sockets, three channels & ZMQ REQ/REP and PUB/SUB \\
Envelope & \code{E3-PDU}, same procedure names & \code{E3-PDU} & JSON \code{common\_headers} \\
Payload bound & 16\,MiB with fragmentation & 32\,KiB, no fragmentation & out-of-band shared memory \\
Control outcome & response with a service-model body & acknowledgement only & JSON response \\
What an existing dApp needs & the published schemas & an envelope adapter; E3DP serves the reference client's transport, but its shipped schema must be regenerated & a new transport layer \\
\bottomrule
\end{tabularx}
\end{table}

\textbf{Transport security}, stated plainly: E3 carries no TLS, DTLS, or token authentication in this release. Both SCTP carriers bind to loopback by default and admit every peer that reaches the bound address unless a literal peer list is configured. A management association carries the full authority of the service model, including load, unload, configuration, and RAN control, so an operator must treat the E3AP port exactly like the local management socket and place an external boundary (IPsec, VPN, network policy) in front of any off-host exposure.

\textbf{Telemetry} (Table~\ref{tab:streams}) differs from prior exports in two properties. \emph{Rate discipline}: per-RNTI floors, per-subscription thinning, and a non-zero-by-default per-stream hard floor are enforced before any capture work, and an idle stream costs one atomic load per event, so subscribing a recorder cannot destabilize the cell; full-rate capture is an explicit operator opt-in. \emph{Completeness}: \code{pusch.grid} carries raw pre-equalization resource elements with every parameter needed to re-run the receive chain (HARQ and RV, scrambling, LDPC base graph, MCS, DM-RS layout, DC position, UCI), the substrate for replay-verified datasets.

\textbf{Management} is one surface. It reports inventory, metrics and metric schemas, incidents, health, served-cell information, and quiet-reservation queries. It drives asynchronous lifecycle jobs, typed configuration against per-package schemas, model operations, and a bounded set of RAN controls (UE release to idle, handover, RU gain) that the host's own authority object admits and defers to the owning component. The same operations work identically over the local socket and SCTP, from the C++ \code{dappctl}, the Python client, or the MCP server of Section~\ref{sec:using}.

\begin{table}[t]
\caption{Telemetry streams and their default hard floors.}
\label{tab:streams}
\centering
\scriptsize
\begin{tabularx}{\columnwidth}{@{}lYl@{}}
\toprule
\textbf{Stream} & \textbf{Content} & \textbf{Floor} \\
\midrule
\code{pusch.symbols} & Equalized symbols, SINR, EVM, EPRE, RSRP, TA, CFO, grant shape & 10\,ms \\
\code{pusch.crc} & Per-transport-block CRC verdict & 1\,ms \\
\code{srs.channel} & Per-subcarrier LS SRS response per rx/tx port pair & 5\,ms \\
\code{pusch.dmrs\_channel} & DM-RS channel estimate per port and layer & 5\,ms \\
\code{pusch.grid} & Raw pre-EQ REs plus all parameters for offline re-decode & 20\,ms \\
\code{gnb.metrics\_json} & The gNB JSON metrics report, byte for byte & report period \\
\code{classc.spectrum} & \ClassC{} snapshot: full uplink grid tensor and expected-use plan & producer rate \\
\code{classc.srs\_isac} & \ClassC{} snapshot: SRS channel-estimate tensor & producer rate \\
\bottomrule
\end{tabularx}
\end{table}

\section{Where the Code Lives}
\label{sec:code}

\begin{figure*}[t]
  \centering
  \resizebox{0.98\textwidth}{!}{
\begin{tikzpicture}[x=1cm, y=1cm, >=Latex,
  repo/.style={draw=#1!85, very thick, fill=#1!6, rounded corners=3pt, inner sep=6pt, align=left, font=\scriptsize\sffamily, text width=47mm, minimum height=42mm},
  seam/.style={draw=ocuE!70, fill=ocuE!8, rounded corners=3pt, inner sep=5pt, align=center, font=\scriptsize\sffamily},
  arr/.style={->, thick, #1!85},
  lbl/.style={font=\tiny\sffamily, text=#1!80!black, fill=white, inner sep=1pt},
]
\node[repo=ocuH] (plat) at (2.8,0) {%
\textbf{\code{ocudu-dapp-platform}} (clone as \code{ocudu/})\\[1pt]
the gNB with the embedded dApp runtime and E3 agent\\[2pt]
\code{include/ocudu/dapp/} the frozen ABI (\code{abi/v1}), the 6 use-case interfaces (\code{use\_cases/v1}), the process contracts (\code{process/v1})\\
\code{lib/dapp/} runtime, E3 agent and codecs, IPC\\
\code{lib/phy/upper/dapp/}, \code{lib/scheduler/dapp/} the Class A and B seams\\
\code{apps/} dApp service, \code{dappctl}, packaging, preflight\\
\code{schemas/e3/}, \code{python/ocudu\_e3/}, \code{docs/dapp/}};
\node[repo=ocuE] (sdk) at (9.4,0) {%
\textbf{\code{ocudu-dapp-sdk}}\\[1pt]
the authoring kit; builds against the installed platform package only\\[2pt]
\code{reference/} 16 reference packages (CE, EQ, receiver, scheduler, spectrum, SRS-ISAC; CPU and CUDA), supervised workers, portable E3 consumers\\
\code{examples/} \code{vendor\_min}, Python Class C over E3, telemetry observers, sensing, MCP server\\
\code{tools/} certify, scaffold, validate; \code{docs/} authoring guides};
\node[repo=ocuA] (ven) at (16.0,0) {%
\textbf{a vendor repository} (yours)\\[1pt]
scaffolded by the SDK; \code{find\_package(OCUDUDAppSDK)}; own CI, own cadence, own models\\[2pt]
output: shared object $+$ signed manifest $+$ SPDX SBOM\\[2pt]
admitted by the operator's trust policy, named in the gNB YAML, driven over E3};
\node[repo=ocuC, text width=108mm, minimum height=0mm] (qs) at (9.4,-3.05) {%
\textbf{\code{ocudu-dapp-quickstart}} (clone as \code{ocudu-docker/}): 1 Dockerfile builds both repositories and runs every release gate; CPU and CUDA images with radio support (UHD, DPDK); \code{docker-compose.yml} with a loopback E3 agent so every tutorial runs without hardware; the tutorial ladder and documentation map};
\node[seam, text width=172mm] (seam) at (9.4,-4.55) {\textbf{the seam everything builds against:} installed CMake package \code{OCUDUDAppSDK}, frozen C ABI v1 with a checked-in layout fingerprint, 6 versioned use-case interfaces, published E3 schemas. No repository links against another; packages meet only at runtime, through the host.};
\draw[arr=ocuC] (sdk.west) -- (plat.east); \draw[-, ocuC!60, thin] (6.1,1.85) -- (6.1,0.15); \node[lbl=ocuC] at (6.1,2.0) {installed headers only};
\draw[arr=ocuA] (ven.west) -- (sdk.east); \draw[-, ocuA!60, thin] (12.7,1.85) -- (12.7,0.15); \node[lbl=ocuA] at (12.7,2.0) {scaffold, certify};
\draw[arr=ocuE] (plat.south) -- (plat.south |- seam.north);
\draw[arr=ocuE] (ven.south) -- (ven.south |- seam.north);
\draw[arr=ocuE] (qs.south) -- (qs.south |- seam.north);
\end{tikzpicture}}
  \caption{The three public repositories and the seam between them. The platform repository owns the interfaces, the runtime, and the E3 plane; the SDK and every vendor repository build against the installed package alone; the quickstart builds all of it into a container with a loopback E3 agent.}
  \label{fig:repos}
\end{figure*}
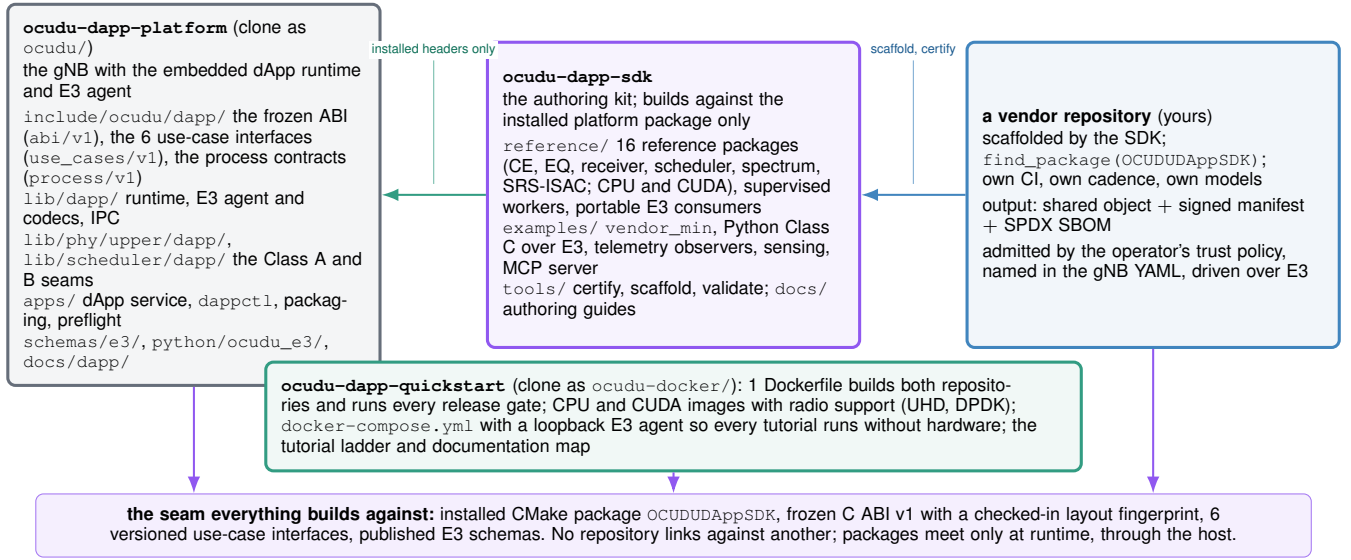

The platform is published as three repositories under the \ocudu{} AI-RAN Working Group~2 (Fig.~\ref{fig:repos}), cloned side by side with the platform as \path{ocudu/} and the quickstart as \path{ocudu-docker/}. The working group develops them in the open. This release is a preview, so others can test the interfaces, propose use cases, and vet the references before the runtime is upstreamed into the \ocudu{} mainline. Table~\ref{tab:entry} maps each mechanism in this paper to its entry point, so a reader can go from a sentence here to the code that implements it.

\textbf{\code{ocudu-dapp-platform}} is the gNB with the embedded runtime. The public ABI is \path{include/ocudu/dapp/abi/v1/} (module, host API, memory, execution, management, metric catalog), the six use-case interfaces are \path{include/ocudu/dapp/use_cases/v1/} together with \path{coordination.h} for the quiet-reservation intent, and the out-of-process contracts (shared ring layout, CUDA export pool, SEQPACKET control protocol) reside in \path{include/ocudu/dapp/process/v1/}. The runtime under \path{lib/dapp/runtime/} holds the package catalog and manifest, the artifact verifier and signature checks, the disposable preflight, the module loader, the native hook registry and slots, the instance manager and lifecycle, the job dispatcher, the circuit breaker, the quiescence tracker, and the \ClassC{} process supervisor. The E3 plane under \path{lib/dapp/e3/} holds the embedded agent, the management service, the APER and FlatBuffers codecs, the E3AP and E3DP SCTP servers, the local credentialed transport, the portable data store, and the telemetry publisher; \path{lib/dapp/ipc/} holds the shared SPSC ring, the SEQPACKET channel, and the \ClassC{} publication routes. The seams into the PHY and MAC are \path{lib/phy/upper/dapp/} and \path{lib/scheduler/dapp/}. The gNB-side service and YAML configuration are \path{apps/services/dapp/}. The operator tools \code{dappctl}, \code{dapp\_package}, and \code{dapp\_preflight} are under \path{apps/tools/}. The schemas are \path{schemas/e3/}, the Python client is \path{python/ocudu_e3/}, and \path{docs/dapp/} holds the thirteen tutorials, the glossary, and the reference documents.

\textbf{\code{ocudu-dapp-sdk}} is the authoring kit. It consumes the installed \code{OCUDUDAppSDK} CMake package and never copies host headers. \path{reference/} holds sixteen reference packages (channel estimator, equalizer, receiver, scheduler, spectrum, SRS-ISAC, in CPU and CUDA variants), the supervised \path{process/} workers, and the \path{portable/} E3 consumers; \path{examples/} holds the copy-me commercial \ClassA{} layout \code{vendor\_min}, a Python \ClassC{} dApp over E3, Python telemetry observers, spectrum and ISAC sensing over E3AP, and the MCP server. \path{tools/} holds the certifier, the scaffolds for every class, and the validation ladder. \path{docs/} holds the per-class authoring guides.

\textbf{\code{ocudu-dapp-quickstart}} is a Dockerfile that builds both repositories from their current commits, runs every release gate inside the build, and produces CPU and CUDA images with radio support (UHD, DPDK) baked in, plus a Compose file with a loopback E3 agent so every tutorial command runs without hardware.

\begin{table}[t]
\caption{Where each mechanism lives (paths relative to the platform repository unless marked SDK).}
\label{tab:entry}
\centering
\scriptsize
\begin{tabularx}{\columnwidth}{@{}L{2.55cm}Y@{}}
\toprule
\textbf{Mechanism} & \textbf{Entry point} \\
\midrule
Frozen C ABI v1 & \path{include/ocudu/dapp/abi/v1/module.h}, \path{host_api.h}, \path{memory.h}, \path{execution.h}; fingerprint gate in the SDK CI \\
Use-case interfaces & \path{include/ocudu/dapp/use_cases/v1/} (\path{spectrum.h}, \path{srs_isac.h}, \path{channel_estimator.h}, \path{equalizer.h}, \path{receiver.h}, \path{scheduler.h}, \path{coordination.h}, \path{pusch_metadata.h}) \\
\ClassA{} seam & \path{lib/phy/upper/dapp/}; hook slots in \path{lib/dapp/runtime/native_hook_*.cpp} \\
\ClassB{} seam & \path{lib/scheduler/dapp/}; intent validation and allow flags in the scheduler policy \\
\ClassC{} native and process & \path{lib/dapp/ipc/publication_route.cpp}, \path{shared_spsc_ring.cpp}, \path{seqpacket_channel.cpp}; \path{lib/dapp/runtime/class_c_process_supervisor.cpp} \\
Admission and trust & \path{lib/dapp/runtime/package_catalog.cpp}, \path{package_manifest.cpp}, \path{package_signature.cpp}, \path{artifact_verifier.cpp}, \path{native_preflight.cpp}; \path{docs/dapp/security_model.md} \\
Lifecycle and jobs & \path{lib/dapp/runtime/instance_manager*.cpp}, \path{lifecycle.cpp}, \path{job_dispatcher.cpp}, \path{quiescence_tracker.cpp} \\
Safety & \path{circuit_breaker.cpp}, \path{guarded_call.h}, \path{incident_store.cpp} \\
E3 agent and codecs & \path{lib/dapp/e3/embedded_agent.cpp}, \path{management_service.cpp}, \path{aper_codec.cpp}, \path{e3ap_codec.cpp}, \path{fbs_codec.cpp} \\
Carriers & \path{lib/dapp/e3/local_connector.cpp}, \path{e3ap_sctp_server.cpp}, \path{e3dp_sctp_server.cpp}; \path{docs/dapp/e3_wire.md} \\
Telemetry & \path{lib/dapp/e3/telemetry_publisher.cpp}, \path{portable_data_store.cpp} \\
Schemas & \path{schemas/e3/} (ASN.1 management, telemetry, E3AP; FlatBuffers data profile) \\
YAML and service & \path{apps/services/dapp/}; reference in \path{docs/dapp/yaml_configuration.md} \\
Tools & \path{apps/tools/dappctl}, \code{dapp\_package}, \code{dapp\_preflight}; \path{python/ocudu_e3/} \\
Glossary, limitations & \path{docs/dapp/glossary.md}, \path{docs/dapp/known_limitations.md} \\
References (SDK) & \path{reference/}: \code{channel\_estimator}, \code{equalizer}, \code{receiver}, \code{scheduler}, \code{spectrum}, \code{srs\_isac} (each with a \code{\_cuda} variant), \code{process}, \code{portable} \\
Certify and scaffold (SDK) & \path{tools/}: \path{certify_class_a_cuda.cu}, \path{scaffold_class_a_cuda.py}, \path{scaffold_class_bc_cpu.py}, \path{dapp_validate.py} \\
MCP server (SDK) & \path{examples/e3_mcp_server/}; guide in \path{docs/mcp_companion.md} \\
\bottomrule
\end{tabularx}
\end{table}

\section{Working With the Platform}
\label{sec:using}

\textbf{The first session.} The zero-hardware path builds the platform into a container and points every later command at a loopback E3 agent. A green image is a validated platform, because every release gate runs inside the build.

\begin{lstlisting}
# 1. Clone the three public repositories side by side
WG2=https://gitlab.com/ocudu/work_groups/wg2_ai_ran
git clone $WG2/ocudu-dapp-platform.git ocudu
git clone $WG2/ocudu-dapp-sdk.git
git clone $WG2/ocudu-dapp-quickstart.git ocudu-docker
# 2. Build the release images (host, SDK, every gate)
ocudu-docker/scripts/build.sh
# 3. Start the loopback E3 agent and look around
docker compose -f ocudu-docker/docker-compose.yml up e3-agent
python3 -m ocudu_e3 --port 36423 catalog
python3 -m ocudu_e3 --port 36423 subscribe pusch.crc --maximum 5
python3 -m ocudu_e3 --port 36423 inventory
\end{lstlisting}

\textbf{The tutorial ladder.} The documentation is organized as thirteen tutorials (Table~\ref{tab:tutorials}) that take a reader from an empty vendor repository to an LLM agent tuning a live cell. The first five need no signing keys, and all of them run against the loopback agent.

\begin{table}[t]
\caption{The tutorial ladder (\code{docs/dapp/} in the platform, \code{docs/} in the SDK).}
\label{tab:tutorials}
\centering
\scriptsize
\begin{tabularx}{\columnwidth}{@{}clY@{}}
\toprule
\textbf{\#} & \textbf{Tutorial} & \textbf{Covers} \\
\midrule
0 & Zero-hardware testbed & build the images; loopback E3 agent; run everything radio-free \\
1 & New vendor dApp repo (SDK) & scaffold, build, certify against the installed SDK \\
2 & Install and load from the gNB YAML & catalog, trust policy, initial instances, admission \\
3 & Operate with \code{dappctl} & lifecycle, A/B version swap, deactivate and unload, breakers \\
4 & Monitor and tune over MCP (SDK) & ask the gNB questions; KPI-driven LLM tuning loops \\
5 & Signing, SBOMs, keys & what is signed, the SBOM, trust policy, rotation, revocation \\
6 & Exposing dApp KPIs over E3 & declare metrics; read them from any client \\
7 & First dApp: standalone \ClassC{} & a Python out-of-process dApp: subscribe, decode, act back \\
8 & Telemetry and data capture & all eight streams, floors, HDF5 capture, offline PUSCH replay \\
9 & Neural-model operations & stage, validate, warm, activate, rollback; live A/B \\
10 & \ClassB{} scheduler control & the 100\,\us{} contract, allow flags, evaluation-first, sensing to avoidance \\
11 & Troubleshooting runbook & symptom, cause, fix, from real failure modes \\
12 & Capstone & capture, train, package, shadow, activate, monitor, rollback \\
\bottomrule
\end{tabularx}
\end{table}

\textbf{What a module looks like.} Every in-process dApp, whatever its class, exports exactly one symbol, \code{ocudu\_dapp\_query\_v1}, which fills a module table: identity, the interfaces it implements, and the non-real-time lifecycle callbacks \code{create}, \code{configure}, \code{prepare}, \code{warmup}, \code{start}, \code{stop}, \code{quiesce}, \code{destroy}. Each interface supplies one real-time entry point. Listing~\ref{lst:classb} shows a complete, if minimal, \ClassB{} module condensed from the SDK's reference scheduler: it echoes the conventional priority for every candidate, which is what makes it a baseline replica, and lowers the MCS of a UE whose recent block error rate is high, but only when the operator has granted MCS authority. Every structure begins with its size and ABI version, no pointer to a host object crosses the boundary, and the module reads only the size it was given. The SDK's helper layer fills the module table and the worker context from a few lines of C++, so a vendor writes the algorithm, not the plumbing.

\begin{lstlisting}[language=C, caption={A minimal \ClassB{} module against \code{use\_cases/v1/scheduler.h}.}, label={lst:classb}]
#include "ocudu/dapp/use_cases/v1/scheduler.h"
#define SCH(x) OCUDU_DAPP_SCHEDULER_##x      /* shorten the ABI names */

static ocudu_dapp_status_v1
invoke(void* ctx, const ocudu_dapp_scheduler_input_v1* in,
       ocudu_dapp_scheduler_output_v1* out)
{
  /* in->invocation.deadline_monotonic_ns bounds this call */
  uint32_t n = 0;
  for (uint32_t i = 0; i < in->nof_candidates; ++i) {
    const ocudu_dapp_scheduler_candidate_v1* c = &in->candidates[i];
    ocudu_dapp_scheduler_intent_v1* it = &out->intents[n++];
    it->struct_size  = sizeof(*it);
    it->candidate_id = c->candidate_id;
    it->flags        = SCH(INTENT_PRIORITY_V1);
    it->priority     = c->conventional_priority; /* baseline replica */
    if ((in->allowed_output_flags & SCH(ALLOW_MCS_V1)) &&
        (c->flags & SCH(CANDIDATE_BLER_VALID_V1)) &&
        c->bler > 0.2f && c->recommended_mcs > 0) {
      it->flags |= SCH(INTENT_MCS_V1);
      it->recommended_mcs = (int16_t)(c->recommended_mcs - 1);
    }
  }
  out->nof_intents         = n;
  out->decision_generation = in->candidate_generation;
  return OCUDU_DAPP_OK_V1;  /* host validates every field, then commits */
}

/* interface table: worker-context hooks + one real-time entry point */
static const ocudu_dapp_scheduler_interface_v1 api =
    { WORKER_API, invoke, {0} };

/* the only exported symbol; the descriptor names interface
   0x00010006, Class B, in-process placement, per-worker threading */
OCUDU_DAPP_EXPORT_V1 ocudu_dapp_status_v1
ocudu_dapp_query_v1(const ocudu_dapp_host_info_v1* host,
                    ocudu_dapp_module_v1* m)
{
  return populate(host, m, "com.vendor.sched", &descriptor, &api);
}
\end{lstlisting}

\textbf{Building and certifying a dApp.} A vendor scaffolds a project with \code{scaffold\_class\_a\_cuda.py} or \code{scaffold\_class\_bc\_cpu.py}, builds it with \code{find\_package(OCUDUDAppSDK)} against the installed package, and runs \code{certify\_class\_a\_cuda} (for \ClassA{}) or \code{dapp\_validate.py}. The certifier loads the final shared object through the public ABI, runs the complete lifecycle, creates two workers on independent CUDA streams, and drives deterministic vectors from the one-PRB minimum to the package's declared maximum. Two of those vectors matter especially: off-origin placements catch a module that interprets DM-RS coordinates relative to its own allocation, and rank-deficient two-layer references resolve only when the module consumes the joint multi-symbol pilot set. Every caller-owned output sits at a nonzero offset inside a poisoned allocation with prefix, suffix, and inactive-capacity canaries checked after every call. \code{dapp\_package} then emits the manifest and SBOM and signs the bundle with the vendor's key.

\textbf{Loading and driving a dApp.} The operator stages the signed bundle in the catalog directory named in the gNB YAML and drives it over E3. The YAML names manifests only; E3 can select from the catalog but never supply a path.

\begin{lstlisting}
# gnb.yml: the dapp block (production shape)
dapp:
  catalog_root: /opt/ocudu/dapps
  native_preflight_helper: /opt/ocudu/bin/ocudu-dapp-preflight
  package_signature_mode: required        # lab: allow_unsigned
  package_trust_store: /etc/ocudu/dapp-trust.json
  minimum_trust_policy_epoch: 1
  package_manifests: [acme_eq.manifest.json, acme_sched.manifest.json]
  initial_instances:
    - {package_manifest: acme_eq.manifest.json, placement: in_process,
       backend: cuda, target: shadow}      # warmed, zero traffic
  scheduler_evaluation_only: true          # measure before authority
  scheduler_allow_mcs: true                # one allow flag per intent class
  local_e3: {socket_path: /run/ocudu/dapp-management.sock}
  e3ap_sctp: {bind_address: 127.0.0.1, port: 36423}
\end{lstlisting}

\begin{lstlisting}
# walk the lifecycle over the local socket; every step is an async job
ocudu-dappctl inventory
ocudu-dappctl load --package <id> --placement in-process --backend cuda
ocudu-dappctl lifecycle prepare      <instance>   # allocate, upload weights
ocudu-dappctl lifecycle arm          <instance>   # bind hooks, gate closed
ocudu-dappctl lifecycle enter-shadow <instance>   # warm, zero traffic
ocudu-dappctl model  <instance> stage|validate|warm|activate|rollback
ocudu-dappctl config <instance> --set olla_target_bler=f64:0.05 --dry-run
ocudu-dappctl lifecycle activate     <instance>   # one release store
ocudu-dappctl metrics <instance>;  ocudu-dappctl incidents;  ocudu-dappctl health
# the same operations over SCTP from any host
python3 ocudu_dappctl.py --sctp <gnb>:36423 inventory
\end{lstlisting}

Two instances of competing packages can be loaded, one active and one waiting in shadow with its weights already warm, and swapped by activating one and deactivating the other. The KPI comparison then runs across activation windows, which is how equalizer variants were compared over the air (Section~\ref{sec:evidence}).

\textbf{Writing a portable \ClassC{} dApp.} A Python process on any host opens an E3AP association, subscribes to \code{classc.spectrum} or \code{srs.channel}, decodes each indication into a NumPy array, and acts back through management operations. Listing~\ref{lst:python} is the core of the SDK's emitter detector, which was active in the composition run of Section~\ref{sec:evidence}. On a 51-PRB cell it keeps up with the full producer rate, about 120 grids per second at 4\,MB/s, and decodes each grid in about 120\,\us{}. Nothing in the gNB is patched or rebuilt.

\begin{lstlisting}[language=Python, caption={A portable \ClassC{} radiometer over E3AP (\code{examples/e3\_spectrum\_sensing}).}, label={lst:python}]
from ocudu_e3 import E3Client, SubscriptionConfig
from ocudu_e3.classc import decode_spectrum

with E3Client("gnb-host", 36423, dapp_name="radiometer") as client:
    client.setup()                # E3AP handshake; assigns a dApp id
    cfg = SubscriptionConfig("classc.spectrum", drop_policy="dropOldest")
    client.subscribe(["classc.spectrum"], configs=[cfg])
    for ind in client.indications(timeout=1.0):
        rec  = ind.record         # OCC3 header + raw grid bytes
        snap = decode_spectrum(rec.metadata, rec.tensor)
        p_db = 10 * np.log10(snap.prb_power() + 1e-20)  # [symbol, PRB]
        # track a floor only where the plan says "planned empty";
        # report PRB runs above it that persist across slots
        for event in detector.detect(snap):
            print(event, ind.dropped_objects)
\end{lstlisting}

\textbf{Driving the cell from an agent.} The MCP server in the SDK (\path{examples/e3_mcp_server/}) is an E3 client that exposes the interface's KPIs and knobs to LLM agents in the simplest possible form: 28 Model Context Protocol tools over one E3 association, twenty that observe (sector state, per-UE KPIs, traffic history, SRS channel, constellation, telemetry catalog, dApp inventory, health, metrics, configuration, incidents, jobs, quiet reservations) and eight that act (dApp configuration and model operations, lifecycle, UE release and handover, RF gain). Every state-changing tool dry-runs by default; applying a change is a second, explicit call with \code{dry\_run=false}, and a \code{--read-only} flag removes the acting tools entirely. Listing~\ref{lst:mcp} condenses a session recorded against the reference cell on 26 August 2026. The agent needed no protocol knowledge, and the platform's guardrails (typed validation, generations, allow flags, host-arbitrated RAN control) bounded every call: out-of-range values were refused by the dApp's own validator and a stale generation by the host, both exercised live.

\begin{lstlisting}[caption={An LLM agent session over MCP, condensed (reference cell, 26 August 2026).}, label={lst:mcp}]
user > How is the sector looking?
agent> get_cell_overview()
       -> 4 RNTIs in 10 s; UL SINR -10..9.7 dB; UL BLER mean 0.52, worst 1.0
agent> get_kpis(45); list_ues()
       -> 0x4711: SNR 22 dB, CQI 15, UL MCS 15-18, BLER 10-12% on 4000 TBs
       -> 0x474a, 0x474f, ...: SNR -10 dB, MCS 0, BLER ~100%, a few hundred TBs
agent: one real user sitting on the host OLLA's 10% target; the rest are
       a device failing attach at the cell edge. An RF problem, not scheduling.
user > Can we tighten link adaptation for the good UE?
agent> get_dapp_config(instance=sched, schema=true)   -> 22 keys, all passthrough
agent> set_dapp_config(instance=sched, dry_run=true,
         {olla_mode: contained_baseline, mcs_mode: table_plus_olla,
          olla_target_bler: 0.10, olla_step_db: 0.05, olla_max_offset_db: 6.0})
       -> dryRunSuccess
agent: proposal validated; apply with dry_run=false when you are ready,
       then compare BLER and MCS over a comparable window and keep or revert.
\end{lstlisting}

\section{Seven Use Cases on One Architecture}
\label{sec:walkthroughs}

Each use case below is stated as what runs, which path it exercises, and what the platform provides. All of them ran on the live cell of Section~\ref{sec:evidence} except the continual-learning loop, whose end-to-end campaigns ran on a dummy-RU DU.

\textbf{Spectrum sensing} descends directly from the original dApp demo and runs here at three escalating trust levels. A Python E3 observer subscribes to the \ClassC{} spectrum stream, functionally the Northeastern demo but over published schemas, under rate floors, and with no patched gNB. The same radiometer then runs as a supervised process dApp (\code{ref\_spectrumd}) under heartbeats and seccomp. Finally the worker acts as \emph{authority}, returning a per-PRB interference map and quiet-period intents that appear in the scheduler's avoidance mask within a few slots. One algorithm, three containment levels, no DU change.

\textbf{ISAC} rides the SRS channel the gNB already estimates: \code{srs\_isac\_monitor.py} sketches a range--Doppler map from \code{srs.channel}; the packaged \ClassC{} references compute a bounded channel-impulse-response preview from the same tensors, the CUDA variant on the producer's stream without ever materializing it on the host.

\textbf{Dataset capture} is a first-class product. \code{record\_telemetry.py} writes any stream mix to HDF5; \code{export\_pusch\_burst.py} captures \code{pusch.grid} bursts; \code{ocudu-pusch-replay} re-decodes them offline with the gNB's own receiver and cross-checks the recorded CRC verdict. Every capture carries its own verified label, channel estimates, and exact PHY configuration.

\begin{figure*}[t]
  \centering
  \resizebox{0.78\textwidth}{!}{
\begin{tikzpicture}
  \node[fboxsolid=ocuA, minimum width=3.4cm, minimum height=0.8cm] (live) at (6.5,6.15)
    {live cell --- model $v_n$ active\\(Class A / B / C dApp)};
  \node[fbox=ocuE, minimum width=3.3cm] (cap) at (11.5,4.85)
    {live telemetry over E3\\\code{pusch.grid/symbols} · \code{srs.channel}\\rate-floored, per-RNTI, idle $=$ 1 atomic};
  \node[fbox=ocuH, minimum width=3.3cm] (data) at (12.4,2.35)
    {training data\\streaming subscription or recorded\\corpora; CRC-verified ground truth};
  \node[fbox=ocuC, minimum width=3.4cm] (train) at (9.2,0.7)
    {online tuning process\\a dApp, RIC app, or edge job ---\\streaming updates or batch retrain};
  \node[fbox=ocuC, minimum width=3.2cm] (pkg) at (4.6,0.7)
    {sealed weights $v_{n+1}$\\versioned model artifact};
  \node[fbox=ocuA, minimum width=3.3cm] (stage) at (1.1,2.05)
    {stage · validate · warm\\on the \textbf{shadow} instance\\(zero traffic exposure)};
  \node[fbox=ocuA, minimum width=3.0cm] (act) at (0.85,4.55)
    {atomic activate\\(generation-checked)};

  \draw[farr=ocuE] (live) to[bend left=14] (cap);
  \draw[farr=ocuH] (cap)  to[bend left=14] (data);
  \draw[farr=ocuC] (data) to[bend left=14] (train);
  \draw[farr=ocuC] (train) to[bend left=10] (pkg);
  \draw[farr=ocuA] (pkg)  to[bend left=14] (stage);
  \draw[farr=ocuA] (stage) to[bend left=14] (act);
  \draw[farr=ocuA] (act)  to[bend left=14] (live);

  \node[fnote=ocuE, align=center] at (11.5,6.15)
    {continuous monitoring drives the loop:\\per-model metrics, fallback counters, incidents, KPIs};

  \draw[fdash=ocuR] (live.south) to[bend right=22] (act.east);
  \node[fnote=ocuR, align=center] at (3.6,4.55) {regression $\Rightarrow$ rollback\\(previous model retained)};

  \node[flabel=ocuE, align=center] at (6.6,3.4)
    {continual AI-RAN model lifecycle\\--- one E3 surface end to end ---};
  \node[fnote=ocuH, align=center] at (6.6,2.7)
    {every step is a typed, transaction-id'd,\\dry-runnable management operation};
\end{tikzpicture}}
  \caption{The continual-learning loop. Rate-floored telemetry streams into an online tuning process; sealed model artifacts return over E3, staged and warmed on a shadow instance, and activated atomically; monitoring drives rollback or the next round. One E3 surface carries the whole loop.}
  \label{fig:mlops}
\end{figure*}
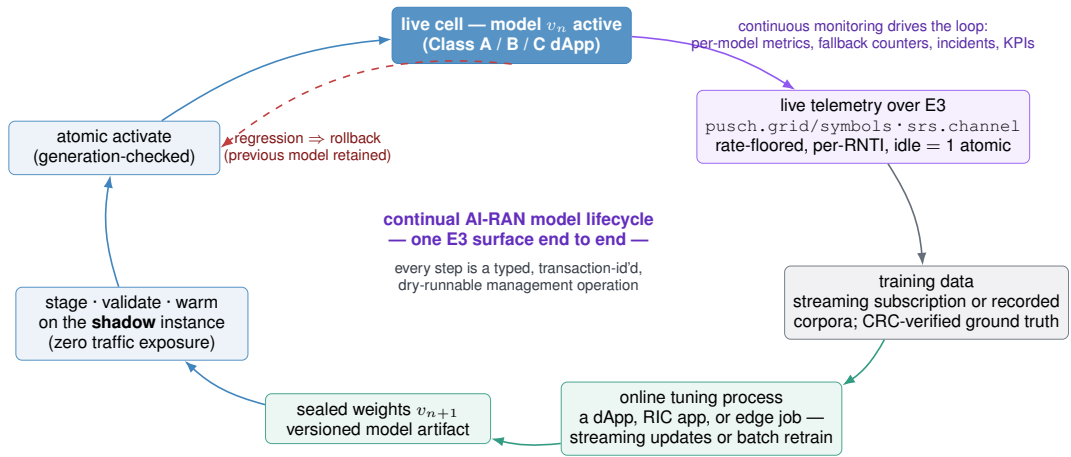

\textbf{The continual-learning loop} (Fig.~\ref{fig:mlops}) assembles those pieces. Telemetry feeds an online tuning process, which may run as a dApp, a RIC application, or an edge job. That process returns a sealed model artifact over the same E3 surface. The artifact is staged, validated, and warmed on a shadow instance with zero exposure, then activated atomically under a generation check. Per-model metrics and incidents monitor it; degradation triggers a one-operation rollback; drift re-enters at capture. Every step is a typed, transaction-id'd, dry-runnable operation, so the loop can be driven by a script, a CI pipeline, or an agent.

\textbf{The inline neural receiver} is the flagship \ClassA{} application. The SDK's CUDA references implement the exact contracts a neural receiver adopts at the three depths. They also carry the model machinery a learned implementation needs: prepare-time weight upload, per-worker lane-local history, and two bounded device weight banks with completion-fence discipline, so an inactive bank is never freed while a stream still reads it. The neural equalizer that ran on the cell is \emph{not} in the SDK. It is a package built out of source tree, against the public SDK alone, by a separate team, which is the seam this paper claims, demonstrated.

\textbf{The inline learned scheduler.} A learned policy replaces stages behind the same intent schema as the baseline replica and inherits the same evaluation-first path, the deadline, the post-equalization noise from \ClassA{}, and the interference context from \ClassC{}, so it can be interference-aware on day one.

\textbf{Agent-driven operations.} Because every operation is a typed request defined by a published schema, exposing the whole surface to an LLM agent takes only a thin adapter. An agent can run a standing loop, read KPIs and incidents every few minutes, reason about trends, and adjust cell parameters or dApp configuration from a single natural-language prompt, with the same guardrails that make the loop safe for a script.

\section{The Vendor Ecosystem: Build, Load, Compose}
\label{sec:ecosystem}

\begin{figure*}[t]
  \centering
  \resizebox{0.98\textwidth}{!}{
\begin{tikzpicture}
  \node[flane=ocuH, fit={(0,0.2) (4.6,6.3)}, inner sep=0pt] {};
  \node[flabel=ocuH, anchor=north west] at (0.1,6.25) {independent vendor repositories};
  \node[fbox=ocuA, minimum width=4.1cm] (v1) at (2.3,5.1)
    {\textbf{Vendor A} --- neural receiver\\Class A CUDA, own models\\\code{find\_package(OCUDUDapp)}};
  \node[fbox=ocuB, minimum width=4.1cm] (v2) at (2.3,3.7)
    {\textbf{Vendor B} --- scheduler policy\\Class B intents pipeline};
  \node[fbox=ocuC, minimum width=4.1cm] (v3) at (2.3,2.3)
    {\textbf{Vendor C} --- sensing / ISAC\\Class C suite $+$ E3 apps};
  \node[fnote=ocuH, align=left, anchor=west] at (0.25,1.1)
    {each repo: own CI, own release cadence,\\scaffold-generated start, offline\\certification --- \emph{no OCUDU source needed}};

  \node[flane=ocuE, fit={(5.2,0.2) (9.2,6.3)}, inner sep=0pt] {};
  \node[flabel=ocuE, anchor=north west] at (5.3,6.25) {signed packages};
  \node[fbox=ocuE, minimum width=3.5cm] (p1) at (7.2,5.1)
    {\code{nrx.so} · manifest-v1 ·\\SPDX SBOM · detached sig};
  \node[fbox=ocuE, minimum width=3.5cm] (p2) at (7.2,3.7)
    {\code{sched.so} · manifest ·\\SBOM · sig};
  \node[fbox=ocuE, minimum width=3.5cm] (p3) at (7.2,2.3)
    {\code{sense.so} · manifest ·\\SBOM · sig};
  \node[fnote=ocuE, align=left, anchor=west] at (5.4,1.1)
    {immutable, hash-bound artifacts;\\SBOM identity independently\\checked at admission};
  \draw[farr=ocuA] (v1.east) -- (p1.west);
  \draw[farr=ocuB] (v2.east) -- (p2.west);
  \draw[farr=ocuC] (v3.east) -- (p3.west);

  \node[flane=ocuR, fit={(9.8,0.2) (13.6,6.3)}, inner sep=0pt] {};
  \node[flabel=ocuR!80!black, anchor=north west] at (9.9,6.25) {operator / SI};
  \node[fbox=ocuR, minimum width=3.3cm] (trust) at (11.7,5.05)
    {\textbf{offline trust policy}\\per-vendor keys · epochs ·\\revocation · sequence floors};
  \node[fbox=ocuR, minimum width=3.3cm] (cat) at (11.7,3.55)
    {\textbf{catalog}\\\code{/opt/ocudu/dapps}\\YAML names manifests only};
  \node[fbox=ocuR, minimum width=3.3cm] (knobs) at (11.7,1.9)
    {\textbf{tuning knobs}\\allow flags · budgets · floors ·\\placement · shadow-first A/B};
  \draw[farr=ocuE] (p1.east) .. controls +(0.9,-0.4) .. (cat.west);
  \draw[farr=ocuE] (p2.east) -- (cat.west);
  \draw[farr=ocuE] (p3.east) .. controls +(0.9,0.4) .. (cat.west);

  \node[flane=ocuH, fit={(14.2,0.2) (18.4,6.3)}, inner sep=0pt] {};
  \node[flabel=ocuH, anchor=north west] at (14.3,6.25) {one OCUDU DU, at runtime};
  \node[fboxsolid=ocuA, minimum width=3.7cm, minimum height=0.62cm] (sa) at (16.3,5.15)
    {A: Vendor A neural RX};
  \node[fboxsolid=ocuB, minimum width=3.7cm, minimum height=0.62cm] (sb) at (16.3,4.25)
    {B: Vendor B scheduler};
  \node[fboxsolid=ocuC, minimum width=3.7cm, minimum height=0.62cm] (sc) at (16.3,3.35)
    {C: Vendor C sensing ($\times N$)};
  \node[fnote=ocuH, align=center] at (16.3,2.45)
    {loaded / swapped / unloaded live\\over E3 --- \textbf{no DU rebuild, no restart}};
  \node[fnote=ocuR, align=center] at (16.3,1.35)
    {host arbitrates: no cross-vendor linkage;\\cooperation only via host-owned caches;\\conventional fallback always armed};
  \draw[farr=ocuR] (cat.east) .. controls +(0.9,0.3) .. (14.2,4.25)
    node[pos=0.4, above=2pt, fnote=ocuR] {admit $+$\\lifecycle};
  \draw[fdash=ocuR] (trust.east) .. controls +(1.2,-0.2) .. (14.2,4.9);
\end{tikzpicture}}
  \caption{The ecosystem the infrastructure enables. Vendors build signed packages in their own repositories against the installed SDK; operators and integrators hold the trust policy, compose packages from several vendors in one catalog, and load, tune, swap, and unload them on a live DU: no rebuild, no cross-vendor linkage, host arbitration throughout.}
  \label{fig:ecosystem}
\end{figure*}
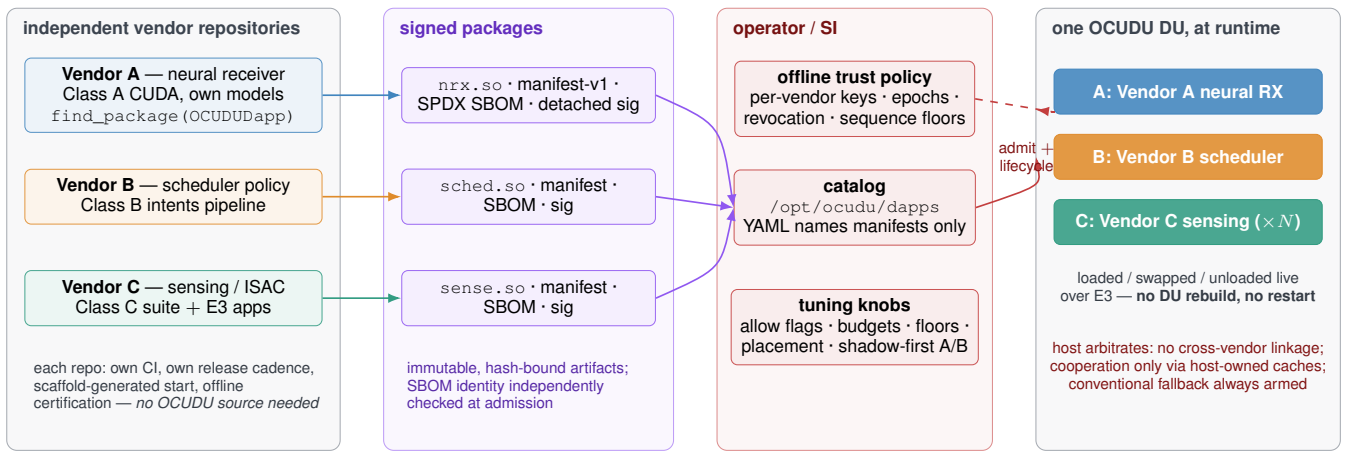

Everything in Section~\ref{sec:walkthroughs} was one team's applications. The same infrastructure supports teams that have never met (Fig.~\ref{fig:ecosystem}) through one clean seam: the host repository owns the six interfaces, the E3 plane, and the runtime; the installed SDK package, the frozen ABI, and the published schemas are the only things anyone builds against. Every producer of dApps sits above that seam with the same headers, certification tools, and packaging.

\textbf{Vendors: your repository, your cadence.} A dApp lives in the vendor's own repository, built against the installed SDK with \code{find\_package}: no host source tree, no fork, no patch queue. The fingerprint-verified ABI and the published compatibility policy decouple the vendor's release cadence from the host's. Offline certification and the validation ladder give the vendor a proof to ship beside the artifact, an immutable, hash-bound bundle of shared object, manifest, SPDX SBOM, and detached signature.

\textbf{Operators and integrators: compose, tune, optimize.} The operator holds the other half of the trust relationship: the offline trust policy of Section~\ref{sec:lifecycle}. Composition is a catalog decision, signed bundles from several vendors named in YAML and driven over E3, and it is safe by construction because packages cannot link against or call each other, cooperate only through host-owned caches, and every authority is individually granted. The operator therefore tunes the combination: allow flags decide how much scheduling authority a policy gets, budgets and floors decide how much real-time and telemetry cost a sensing suite may impose, placement decides its containment, and shadow-staged candidates with per-instance metrics make A/B between competing packages routine.

\textbf{A vendor-neutral home, and a path to standards.} The code lives under the \ocudu{} AI-RAN Working Group~2 as a preview release, and the working group is the venue in which it is being vetted, extended with new use cases, and prepared for upstreaming into the \ocudu{} mainline. Interfaces defined in code, exercised by shipping references, and stress-tested by vendors interworking through them let technical consensus form in the project, at the speed of working software. What then goes to the SDOs, O-RAN's E3 work~\cite{oran_ngrg} and ultimately 3GPP, is a proven artifact. The \ClassA{} and \ClassB{} C ABIs are deliberately portable beyond this codebase, and the E3 end already admits any stack that speaks the schemas.

\section{The Latency Architecture and Safety}
\label{sec:safety}

A slot is 500\,\us{} at 30\,kHz subcarrier spacing, and at five-cell CUDA L1 scale roughly 100\,\us{} of host jitter already produces fronthaul lateness, so the platform must add nothing measurable to the radio's steady state. The design guarantees this by construction (Fig.~\ref{fig:latency}) in two moves. \emph{First}, every steady-state mechanism on a real-time thread costs nanoseconds to single-digit microseconds: an atomic gate when idle, a CAS for a rate floor, a lock-free cache read, a try-lock publication with an eventfd write. \emph{Second}, work is placed by thread, not by module: the real-time lane's entire E3 duty is one gated memcpy, while encoding, socket I/O, and lifecycle jobs live on management executors that are allowed to be slow. Five rules make this concrete:
\begin{enumerate}
\item \textbf{Idle is an atomic load.} Every capture site and dApp hook is gated by a single relaxed or acquire atomic. An unconfigured feature has no measurable cost.
\item \textbf{Producers never block.} Try-locks, non-blocking eventfds, and counted drops replace every wait a consumer could induce. A slow observer is its own problem, never the radio's.
\item \textbf{Every residual wait is bounded.} A degraded GPU or wedged consumer costs at most the current slot: fallback fires, a counter increments, no thread stalls.
\item \textbf{No steady-state allocation} on any invocation path: preallocated workspaces, lock-free snapshot pools, builders cleared per encode.
\item \textbf{GPU discipline.} \ClassA{} enqueues on the host's stream and returns. \ClassC{} CUDA export is one device-to-device snapshot into a bounded IPC pool, never a mapping of live L1 memory.
\end{enumerate}

\begin{figure*}[t]
  \centering
  \resizebox{0.98\textwidth}{!}{
\begin{tikzpicture}[x=1.9cm, y=1cm, >=Latex,
  pin/.style={font=\scriptsize\sffamily, text=#1!70!black, align=center},
  lanebox/.style={draw=#1!55, fill=#1!5, rounded corners=4pt, inner sep=0pt},
  lanelbl/.style={font=\scriptsize\sffamily\bfseries, text=#1!85!black, anchor=west},
  item/.style={draw=#1!75, fill=#1!10, rounded corners=2pt, font=\scriptsize\sffamily, inner sep=3pt, anchor=west},
  lanetext/.style={font=\scriptsize\sffamily, text=#1!80!black, anchor=north west, align=left, text width=13.4cm},
]
\draw[line width=0.8pt, ->, draw=ocuH] (-0.1,0) -- (7.3,0);
\foreach \x/\lbl in {0/{1\,ns},1/{10\,ns},2/{100\,ns},3/{1\,$\mu$s},4/{10\,$\mu$s},5/{100\,$\mu$s},6/{1\,ms},7/{10\,ms}}{
  \draw[draw=ocuH] (\x,-0.07) -- (\x,0.07);
  \node[font=\scriptsize\sffamily, text=ocuH, below=1pt] at (\x,-0.07) {\lbl};
}
\node[font=\scriptsize\sffamily\bfseries, text=ocuH, anchor=west] at (-0.1,-0.7) {cost per event, log scale (measured on the GB10 host)};
\draw[dashed, line width=0.7pt, draw=ocuR!75] (5.0,-0.1) -- (5.0,2.75);
\node[pin=ocuR, anchor=south] at (5.0,2.78) {about 100\,$\mu$s of host jitter\\already means fronthaul lateness};
\draw[dashed, line width=0.7pt, draw=ocuH!75] (5.7,-0.1) -- (5.7,2.0);
\node[pin=ocuH, anchor=south] at (5.7,2.03) {1 slot\\(500\,$\mu$s)};
\foreach \x/\y/\col/\lbl in {
    0.5/0.95/ocuC/{idle gate:\\1 atomic load},
    1.05/1.95/ocuC/{rate floor:\\1 CAS},
    1.68/0.95/ocuC/{context cache read:\\48\,ns P50},
    2.46/1.95/ocuB/{Class B direct call:\\0.29\,$\mu$s P99.9},
    3.3/0.95/ocuC/{ring publish, 68\,KB:\\2.3\,$\mu$s P99.9},
    3.95/1.95/ocuE/{E3AP control RT:\\11\,$\mu$s P50}}{
  \fill[\col] (\x,0.13) circle (2.2pt);
  \draw[draw=\col!70, line width=0.5pt] (\x,0.13) -- (\x,\y-0.36);
  \node[pin=\col] at (\x,\y) {\lbl};
}
\fill[ocuA] (4.91,0.13) circle (2.2pt);
\draw[draw=ocuA!70, line width=0.5pt] (4.91,0.13) -- (6.15,0.62);
\node[pin=ocuA, anchor=west] at (6.2,0.75) {receiver kernels, live shape:\\82\,$\mu$s P50 (GPU, async)};
\begin{scope}[yshift=-1.35cm]
\node[lanebox=ocuA, fit={(-0.1,-1.7) (7.3,-0.95)}] {};
\node[lanelbl=ocuA] at (0.0,-1.13) {UL-PHY worker, real-time priority, per cell: the only 4 operations it performs for dApps};
\node[item=ocuA] at (0.0,-1.48) {atomic gate};
\node[item=ocuA] at (1.05,-1.48) {CAS on the rate floor};
\node[item=ocuA] at (2.55,-1.48) {1 memcpy or 1 device-to-device enqueue};
\node[item=ocuA] at (5.05,-1.48) {try-lock $+$ eventfd write};
\node[lanebox=ocuR, fit={(-0.1,-2.92) (7.3,-1.85)}] {};
\node[lanelbl=ocuR] at (0.0,-2.03) {never on that thread};
\node[lanetext=ocuR] at (0.0,-2.2) {allocation, locks that wait, any unbounded wait, ASN.1 or FlatBuffers encoding, log formatting, disk or network I/O; every residual wait is deadline-bounded and costs at most the slot};
\node[lanebox=ocuE, fit={(-0.1,-3.9) (7.3,-3.02)}] {};
\node[lanelbl=ocuE] at (0.0,-3.2) {management executors, normal priority: all the expensive work};
\node[lanetext=ocuE] at (0.0,-3.37) {APER and FlatBuffers encoding, SCTP and socket I/O, lifecycle jobs, store leases, incidents, metrics, model staging};
\end{scope}
\end{tikzpicture}}
  \caption{The latency architecture. Top: every steady-state mechanism on the real-time path on a log cost axis, three orders of magnitude inside the slot budget and far below the jitter that already causes fronthaul lateness. Bottom: work placement by thread. The real-time lane owns four cheap operations; everything expensive runs at normal priority.}
  \label{fig:latency}
\end{figure*}

\textbf{What a deadline means, per class.} The deadlines are enforced differently, and an operator should know which is which. For \ClassB{}, the deadline is enforced at three checkpoints and a late result is skipped, discarded, or rolled back, so the radio never sees it. For \ClassA{}, a failure on the callback clock selects the conventional stage in the same invocation, but a kernel that has been enqueued cannot be cancelled, so a late-but-successful completion is used and recorded rather than replaced; the containment for a slow module is the breaker, which opens after eight consecutive misses, together with the admission gate that keeps out-of-profile grants away from it. Preempting native code mid-invocation cannot be done safely, and this is a deliberate v1 decision: a package whose timing is not yet trusted belongs in \ClassC{} process placement, not inline. For \ClassC{}, there is no producer-side deadline at all.

\begin{figure*}[t]
  \centering
  \resizebox{0.98\textwidth}{!}{
\begin{tikzpicture}
  \node[fbox=ocuH, minimum width=2.55cm, minimum height=1.55cm] (s1) at (1.4,2.7)
    {\textbf{author}\\out-of-tree,\\SDK contracts,\\frozen C ABI};
  \node[fbox=ocuH, minimum width=2.55cm, minimum height=1.55cm] (s2) at (4.5,2.7)
    {\textbf{certify offline}\\\code{certify\_class\_a}\\\code{\_cuda} · L1 vectors ·\\canary-checked outputs};
  \node[fbox=ocuE, minimum width=2.55cm, minimum height=1.55cm] (s3) at (7.6,2.7)
    {\textbf{package}\\manifest-v1 · SPDX\\SBOM · detached sig ·\\offline trust policy};
  \node[fbox=ocuE, minimum width=2.55cm, minimum height=1.55cm] (s4) at (10.7,2.7)
    {\textbf{admit}\\hash/SBOM binding ·\\isolated preflight ·\\declared shape profiles};
  \node[fbox=ocuA, minimum width=2.55cm, minimum height=1.55cm] (s5) at (13.8,2.7)
    {\textbf{run guarded}\\\code{noexcept} boundary ·\\deadlines $+$ breakers ·\\incident records};
  \node[fbox=ocuC, minimum width=2.55cm, minimum height=1.55cm] (s6) at (16.9,2.7)
    {\textbf{isolate (Class C)}\\SEQPACKET $+$ creds ·\\seccomp · rlimits ·\\heartbeats · restarts};
  \foreach \a/\b in {s1/s2, s2/s3, s3/s4, s4/s5, s5/s6}{ \draw[farr=ocuH] (\a) -- (\b); }

  \node[flane=ocuR, fit={(0.15,0.2) (18.2,1.0)}, inner sep=0pt] (rail) {};
  \node[flabel=ocuR] at (9.2,0.6)
    {conventional PHY $+$ scheduler, always armed: a fault, lateness, or unload returns stock behavior, never silence};
  \draw[fdash=ocuR] (s4.south) -- (s4.south |- rail.north) node[midway, right=2pt, fnote=ocuR] {reject};
  \draw[fdash=ocuR] (s5.south) -- (s5.south |- rail.north) node[midway, left=2pt, fnote=ocuR] {fallback $+$ breaker};
  \draw[fdash=ocuR] (s6.south) -- (s6.south |- rail.north) node[midway, left=2pt, fnote=ocuR] {kill $+$ bounded\\restart};
  \node[fnote=ocuH, anchor=north west, align=left] at (0.15,0.05)
    {everything observable over E3: per-reason fallback counters, deadline misses, heartbeat latency, restart and incident records};
\end{tikzpicture}}
  \caption{The safety ladder. Certification and signing happen offline; admission is fail-closed; runtime guards convert faults and lateness into fallback plus breaker; supervised isolation bounds even a misbehaving process. The conventional path remains in place at every rung, and every transition is observable over E3.}
  \label{fig:safety}
\end{figure*}
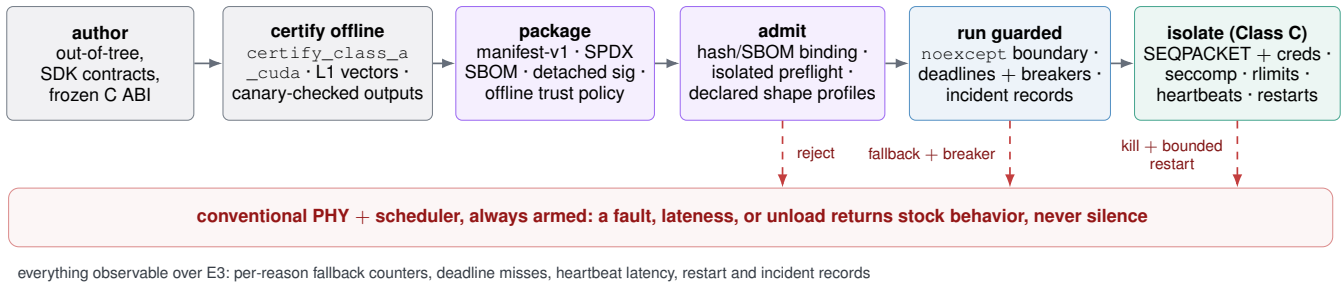

Fig.~\ref{fig:safety} shows the ladder a package climbs and the rail everything falls back to. Admission binds the artifact hash and SBOM identity, runs descriptor discovery in a disposable preflight process, and checks declared shape profiles before any dispatch. Runtime guards catch exceptions at the module boundary and translate them into sanitized incident records, meter deadlines on separate callback and completion clocks, and open per-lane circuit breakers on faults, with the conventional path in place at every rung. Supervised \ClassC{} workers add authenticated \code{SOCK\_SEQPACKET} control with pinned peer credentials, pointer-free results validated at the trust boundary, seccomp deny-lists, resource limits, challenge--response heartbeats, and a backed-off, healthy-uptime-windowed restart budget. Observability closes the loop: per-reason fallback counters, deadline misses with measured GPU durations, heartbeat latencies, restart and incident records, all queryable over the same E3 surface a dApp is managed through. In-process modules remain fully trusted code; preflight and guards reduce the impact of predictable faults and are not a sandbox, and process placement is bounded fault containment, not hostile multi-tenant isolation.

\section{Measured Evidence and Limitations}
\label{sec:evidence}

\begin{figure}[t]
  \centering
  \includegraphics[width=\columnwidth]{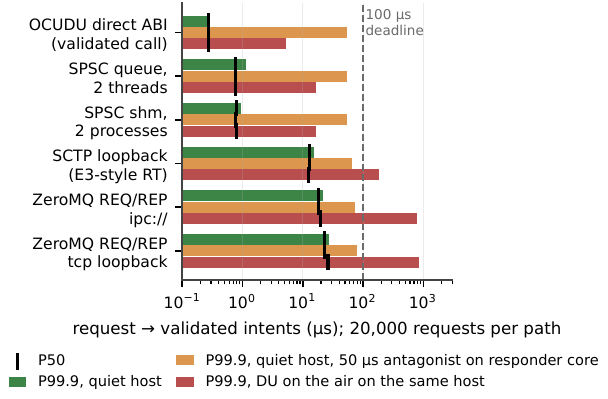}
  \caption{The \ClassB{} contract under load. One 3.2\,KB scheduler request over six boundaries, 20{,}000 requests each; bars are P99.9, ticks P50, the dashed line the admitted deadline. On a quiet host every carrier fits; with the DU on the air only the paths without a wake keep their tails.}
  \label{fig:classb_plot}
\end{figure}

\begin{figure*}[t]
  \centering
  \includegraphics[width=0.9\textwidth]{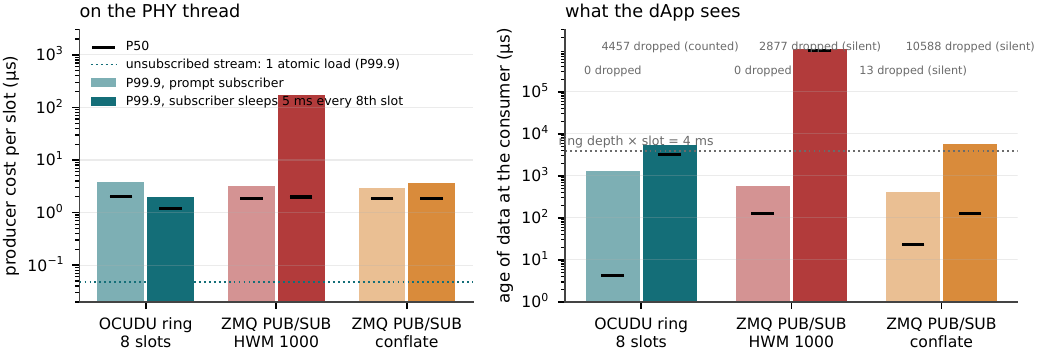}
  \caption{The producer rule, measured. Per-slot producer cost (left) and the age of the data the consumer sees (right) for the \ocudu{} shared ring and ZeroMQ publish/subscribe, with a prompt subscriber and one that sleeps 5\,ms every eighth slot. The ring's producer cost does not move, staleness is bounded by ring depth, and every drop is counted; ZeroMQ queues to its high-water mark or drops silently.}
  \label{fig:producer_plot}
\end{figure*}

\textbf{Testbed.} All live-cell results come from one over-the-air cell: an NVIDIA GB10 (DGX Spark, Arm) host running the CUDA-resident L1, a split-8 USRP B210, band n78 at 3410.1\,MHz, 20\,MHz (51 PRB) TDD at 30\,kHz subcarrier spacing, one transmit and one receive antenna, an Open5GS core, and commercial handsets as UEs, three of them attached during the composition runs. Interface measurements come from the companion interface study~\cite{oshea2026dappstudy}, taken on the same host with the released mechanisms and raw percentiles under two conditions, a quiet host and the cell on the air (Figs.~\ref{fig:classb_plot} and~\ref{fig:producer_plot}). The benchmark suite that produced them is published beside the platform. Tables~\ref{tab:evidence_cost} and~\ref{tab:evidence_system} separate the two kinds of evidence, mechanism costs and system-level runs, and say for each row what a reader needs in order to reproduce it.

\begin{table}[t]
\caption{Mechanism costs and deadline compliance (GB10 host). Runtime rows are the platform's own checkpoints; suite rows are from the companion interface study~\cite{oshea2026dappstudy}, quiet host unless stated.}
\label{tab:evidence_cost}
\centering
\scriptsize
\begin{tabularx}{\columnwidth}{@{}L{2.0cm}L{2.2cm}L{2.4cm}Y@{}}
\toprule
\textbf{Claim} & \textbf{Result} & \textbf{Conditions} & \textbf{Reproduce with} \\
\midrule
\ClassB{} boundary (runtime) & 3.6\,\us{} P99.9, no misses & worst profile, 9{,}000-call development matrix; host without approved CPU controls & platform tests, any x86 or Arm host \\
\ClassB{} direct call (suite) & 0.29\,\us{} P99.9 quiet; 5.2\,\us{} with the cell on air & the same 3.2\,KB request over SCTP loopback: 16\,\us{} quiet, 181\,\us{} on air, 0.35\,\% misses at 100\,\us{} & benchmark suite; on-air rows need the radio \\
\ClassA{} resident receiver & 82\,\us{} P50, 112\,\us{} P99.9 (suite); 92.5 and 105.1\,\us{} (runtime) & live shape, 51 PRB, two layers, 256-QAM, exclusive GPU; 150\,\us{} budget; 273-PRB kernels not yet qualified & certifier on a CUDA host \\
\ClassC{} ring publication & 2.0\,\us{} P50, 2.3\,\us{} P99.9 per 68\,KB slot & producer critical section; full ring is a counted drop; 0.87\,Gb/s sustained across forced worker restarts & benchmark suite, any host \\
E3AP over SCTP & 10\,\us{} one way at 1.5\,KB; 1.4\,ms at 734\,KB; 11\,\us{} control round trip & asn1c APER codec, loopback association, two pinned processes & quickstart loopback agent \\
\bottomrule
\end{tabularx}
\end{table}

\begin{table}[t]
\caption{System-level runs on the live cell.}
\label{tab:evidence_system}
\centering
\scriptsize
\begin{tabularx}{\columnwidth}{@{}L{2.0cm}L{2.5cm}Y@{}}
\toprule
\textbf{Run} & \textbf{Result} & \textbf{Conditions and gaps} \\
\midrule
Composition on one cell & four dApps, all three classes, more than 260{,}000 equalizer invocations, no fallbacks & out-of-tree neural equalizer (A), reference scheduler (B), spectrum and SRS-ISAC (C); three attached handsets. A control window with the same traffic and the dApps unloaded is not yet published \\
Equalizer variants over the air & 7 to 11 percentage points lower first-transmission BLER at MCS 11 to 15 & two neural variants swapped by lifecycle alone, same UEs and traffic; CRC read per RNTI over activation windows. Transport-block counts, window lengths, and confidence intervals are not yet tabulated \\
Operations surface & eight telemetry streams; 28 MCP tools (twenty observe, eight act) & one E3 association; every state-changing tool dry-runs by default; reproducible against the loopback agent \\
\bottomrule
\end{tabularx}
\end{table}

\textbf{Limitations and status.} The gaps are stated as plainly as the features, with what closes each.
\begin{itemize}
\item \emph{\ClassA{} tails.} The reference receiver meets its budget at the live shape on an exclusive GPU, but the 273-PRB kernels complete in about 270\,\us{} against a 150\,\us{} budget and are not yet latency-qualified. Until they are, admission profiles stop at the shapes the certifier has qualified, and a 100\,MHz cell takes the conventional path for the wider grants. Tail qualification is the next admission gate.
\item \emph{\ClassB{} evidence} was gathered on a host without approved CPU controls (no isolated cores or pinned real-time policy), so it is development evidence. The suite shows that on a quiet host every carrier meets the deadline and that only the direct call keeps its tail once the DU is on the air, which is the condition that matters.
\item \emph{Steady-state overhead} is guaranteed by construction and checked by fronthaul-lateness release gates, but the published comparison that the argument of Section~\ref{sec:safety} calls for, the same traffic with the runtime disabled, enabled, and enabled with a full-rate recorder attached, reporting fronthaul lateness, GPU utilization, and per-core load, is still pending. It is the highest-priority measurement before this preview leaves the working group.
\item \emph{Over-the-air comparisons} follow the discipline recorded in the operations cookbook, learned when a sequential comparison once ranked a no-op control above both real arms: interleave and reshuffle the arms, wait for link adaptation to settle after each switch, read CRC counters as deltas within a window from a single RNTI, and accumulate thousands of transport blocks per arm. The block counts, window lengths, and intervals behind the BLER row of Table~\ref{tab:evidence_system} and a control window for the composition run will be published with the qualification report.
\item \emph{Ecosystem.} The multi-vendor machinery is implemented and exercised, yet every package that has run on the cell so far was produced by the authors' organization. The out-of-tree neural equalizer demonstrates the seam, not yet an independent vendor.
\item \emph{Transport security.} E3 carries no transport authentication in this release; the credentialed local socket, a loopback bind, and operator tunnelling are the boundary (Section~\ref{sec:e3}).
\item \emph{Multi-cell.} Telemetry attribution is single-cell today: every stream reports the publisher's default cell id, and host-inline \ClassC{} is not a five-cell configuration. Multi-cell attribution and an asynchronous \ClassC{} prepared-output transaction are roadmap items.
\item \emph{MCP.} The server is an example of the pattern, not a hardened product. Its safety comes from the platform's guardrails.
\end{itemize}

\section{What the Expanded Contract Buys}
\label{sec:compare}

Table~\ref{tab:compare} summarizes the delta against the pioneering frameworks of Section~\ref{sec:prior}, and it should be read as a statement of scope and production maturity: those frameworks set out to prove the tier and succeeded, and much of what follows is machinery one needs only once dApps are trusted enough to matter. Three rows carry the argument. \emph{Authority}: prior control returns were coarse primitives. Here every control path, \ClassA{} outputs, \ClassB{} intents, \ClassC{} maps and quiet requests, E3 RAN control, passes typed validation against operator-granted bounds, and the host remains the oracle. \emph{Producer safety}: prior exports were coupled to L1; here the radio is arithmetically indifferent to its observers. \emph{Operations}: prior dApps were hand-integrated deployments; these are governed artifacts with signatures, SBOMs, generations, dry-runs, incidents, and rollback, which is what makes multi-vendor composition and agent-driven operation possible at all.

\begin{table}[t]
\caption{Prior dApp frameworks (built to prove the tier) versus the \ocudu{} platform (built for production adoption). The delta is scope and maturity, not intent.}
\label{tab:compare}
\centering
\scriptsize
\begin{tabularx}{\columnwidth}{@{}L{1.75cm}L{2.6cm}Y@{}}
\toprule
 & \textbf{Prior (NEU / NVIDIA)} & \textbf{\ocudu{} dApp platform} \\
\midrule
Execution & External process over exported IQ/KPI & External \emph{and} in-process; resident GPU L1 replacement; bounded scheduler hook \\
\midrule
Control & Early primitives (e.g., PRB blacklist) & Typed, validated, allow-flag-bounded intents; granted quiet reservations; RAN control ops \\
\midrule
Producer safety & Export coupled to L1 & Never blocks; idle is one atomic; bounded waits; counted drops \\
\midrule
Fallback & Left to the integration & Conventional path never displaced; breakers; quarantine \\
\midrule
Packaging & Per-deployment integration & Signed manifest, SBOM, offline trust; preflight; ABI fingerprint \\
\midrule
Lifecycle & Manual or scripted & E3 jobs, shadow state, generations, dry-run, rollback \\
\midrule
Model ops & External to the framework & Stage, validate, warm, activate, rollback on a live cell \\
\midrule
Multi-vendor & Vendor-neutral in intent; composition machinery still ahead & Runtime composition under operator trust and allow flags \\
\midrule
Isolation & Container & Supervised process: authenticated IPC, seccomp, heartbeats, bounded restart \\
\midrule
Wire & Pre-standard, still converging & Published ASN.1 APER and FlatBuffers, bounded codecs, versioned schemas; SCTP as first-class carrier \\
\bottomrule
\end{tabularx}
\end{table}

\section{Conclusion and Availability}
\label{sec:conclusion}

\textbf{Artifact availability.} All software is open source under the BSD-3-Clause-Clear license in the \ocudu{} AI-RAN Working Group~2 group, \url{https://gitlab.com/ocudu/work_groups/wg2_ai_ran/}, as three repositories: \code{ocudu-dapp-platform} (the gNB with the dApp runtime, E3 agent, schemas, and Python client), \code{ocudu-dapp-sdk} (reference packages, certifier, scaffolds, examples, and the MCP server), and \code{ocudu-dapp-quickstart} (the reproducible container build, the loopback E3 agent, and the index of tutorials and documentation). The listings in Section~\ref{sec:using} reproduce the E3 surface with no radio, GPU, or signing keys; the mechanism costs of Table~\ref{tab:evidence_cost} reproduce from the published benchmark suite on any host, with the CUDA host and the radio needed only where the table says so; the over-the-air results require a GB10-class host, a USRP or split-7.2 O-RU, and attached UEs, as the tutorials describe. This is a preview release: the working group invites feedback on the interfaces, new use cases that stress them, and independent vetting of the references, all of which feed the upstreaming of the runtime and E3 plane into the \ocudu{} mainline.

\textbf{Outlook.} The ABI and schemas are frozen; the roadmap adds an asynchronous \ClassC{} prepared-output transaction, multi-cell telemetry attribution, and a broadening \ClassB{} feature block as evaluation-mode evidence accumulates. The platform is built to keep growing: new use-case interfaces, new streams, new references, and, by design, vendor dApps the authors will never see, composed by operators the vendors will never meet. Because the interfaces live in a vendor-neutral open-source project, the consensus they accumulate is portable into O-RAN and 3GPP when the time comes to formalize them.

The dApp tier's founders were right about where the opportunity lives, and the observer loop was the correct first step into it. This paper has presented the next step as a working system: an infrastructure on which the same governed artifact can observe the spectrum today, advise the scheduler tomorrow, and replace the receiver the day the evidence supports it, while feeding the datasets that train its own successor, answering to an operations agent, and sharing the cell with dApps from vendors who have never seen its code.

\section*{Acknowledgment}
This work was supported in part by the U.S. DoD OUSD(R\&E) FutureG Office through the National Spectrum Consortium (NSC) Spectrum Forward OTA, including support for contributions to the Linux Foundation's \ocudu{} open-source RAN project.


\end{document}